\documentclass[sigconf,screen]{acmart}

\usepackage{fontawesome} 
\usepackage{microtype}
\newcommand{\toolname}{\textsc{ProbeDebias}}

\usepackage{seqsplit}
\usepackage{enumerate}
\usepackage{ragged2e}
\usepackage{pifont}
\usepackage{stfloats}
\usepackage{amsmath,amsfonts}
\usepackage{graphicx}
\usepackage{textcomp}
\usepackage{xcolor}
\usepackage[normalem]{ulem} 
\usepackage{balance}
\usepackage{hyperref}
\newtheorem{definition}{Definition}
\usepackage{url}
\hypersetup{
    colorlinks=true,
    linkcolor=blue,
    filecolor=blue,      
    urlcolor=blue,
    citecolor=blue,
}
\usepackage{tabularx}  

\usepackage{wrapfig}
\usepackage{rotating}

\usepackage{multirow}
\usepackage[utf8]{inputenc}
\usepackage{graphicx}
\usepackage{textcomp}
\usepackage{color,xcolor}
\usepackage{url}
\usepackage{graphicx}
\usepackage{stmaryrd}
\usepackage{verbatimbox}
\usepackage{enumerate}
\usepackage[shortlabels]{enumitem}
\usepackage{bm}
\usepackage{soul} 
\usepackage{makecell}
\usepackage{booktabs}

\usepackage{multirow}

\usepackage{enumitem}
\usepackage{colortbl}
\usepackage{subcaption}  

\usepackage{subcaption}
\usepackage{booktabs}
\definecolor{bestbg}{gray}{0.9} 

\definecolor{negred}{RGB}{180, 0, 0}

\usepackage{pifont} 

\usepackage{listings}
\usepackage[most]{tcolorbox}
\tcbuselibrary{listingsutf8,breakable}
\definecolor{lsframe}{HTML}{CCCCCC}   
\definecolor{lsbg}{HTML}{F7F7F7}      

\newcommand{\find}[2]{
\begin{tcolorbox}[toprule=0mm,bottomrule=0mm,left=1pt,right=2pt,top=2pt,bottom=2pt,breakable]%
\em #1
\end{tcolorbox}
}

\lstdefinestyle{paperlisting}{
  basicstyle=\ttfamily\scriptsize,
  keywordstyle=\color{blue}\bfseries,
  commentstyle=\itshape\color{teal!70!black},
  stringstyle=\color{magenta!70!black},
  frame=lines,                        
  framerule=0.4pt,
  rulecolor=\color{lsframe},
  backgroundcolor=\color{lsbg},
  xleftmargin=0.6em, xrightmargin=0.6em,
  aboveskip=0.5\baselineskip, belowskip=0.5\baselineskip,
  showstringspaces=false,
  breaklines=true, columns=fullflexible, tabsize=4,
  captionpos=t                        
}

\definecolor{codebg}{RGB}{248,248,248}
\definecolor{codecomment}{RGB}{0,128,0}
\definecolor{codekeyword}{RGB}{0,0,180}
\definecolor{codestring}{RGB}{163,21,21}

\lstdefinestyle{mypython}{
    language=Python,
    basicstyle=\ttfamily\small,
    keywordstyle=\color{codekeyword}\bfseries,
    commentstyle=\color{codecomment}\itshape,
    stringstyle=\color{codestring},
    backgroundcolor=\color{codebg},
    showstringspaces=false,
    breaklines=true,
    frame=single,
    rulecolor=\color{black!15},
    frameround=tttt,
    xleftmargin=4pt,
    xrightmargin=4pt,
    aboveskip=4pt,
    belowskip=4pt
}

\newcommand{\myfinding}[2]{
\begin{center}
\begin{tcolorbox}[colback=gray!20, colframe=gray!100, sharp corners, leftrule={3pt}, rightrule={0pt}, toprule={0pt}, bottomrule={0pt}, left={2pt}, right={2pt}, top={0pt}, bottom={0pt}]
\textbf{\ding{72} Finding {#1}:}
{#2}
\end{tcolorbox}
\end{center}
}

\usepackage{xspace}
\newcommand{\ie}{\textit{i.e.,}\xspace}
\newcommand{\eg}{\textit{e.g.,}\xspace}

\newcommand{\etal}{\textit{et al.}\xspace}

\makeatletter
\newtheoremstyle{plainnoparen}
  {6pt} {6pt}
  {\itshape}
  {}
  {\bfseries}
  {.}
  {.5em}
  {\thmname{#1}~\thmnumber{#2}.\ \thmnote{\itshape #3}} 
\makeatother

\theoremstyle{plainnoparen}

\usepackage{listings}
\usepackage[most]{tcolorbox}
\tcbuselibrary{listingsutf8,breakable}
\definecolor{lsframe}{HTML}{CCCCCC}   
\definecolor{lsbg}{HTML}{F7F7F7}      

\AtBeginDocument{%
  }

\setcopyright{cc}
\setcctype{by}
\acmDOI{10.1145/3832783.3834387}
\acmYear{2026}
\copyrightyear{2026}
\acmISBN{979-8-4007-2882-2/2026/10}
\acmConference[ASE '26]{Proceedings of the 41st IEEE/ACM International Conference on Automated Software Engineering}{October 12--16, 2026}{Munich, Germany}
\acmBooktitle{Proceedings of the 41st IEEE/ACM International Conference on Automated Software Engineering (ASE '26), October 12--16, 2026, Munich, Germany}
\acmSubmissionID{ase26main-p1394-p}
\received{2026-03-26}
\received[accepted]{2026-06-18}

\begin{document}

\title{How Reasoning Shapes Social Bias in LLM-Generated Code?}


\author{Weifeng Sun}
\orcid{0000-0001-6013-1369}
\affiliation{%
  \institution{Singapore Management University}
  \city{Singapore}
  \country{Singapore}
}
\email{wfsun@smu.edu.sg}

\author{Jieke Shi}
\orcid{0000-0002-0799-5018}
\affiliation{%
  \institution{Singapore Management University}
  \city{Singapore}
  \country{Singapore}
}
\email{jiekeshi@smu.edu.sg}

\author{Zhou Yang}
\orcid{0000-0001-5938-1918}
\affiliation{%
  \institution{University of Alberta \\
  Alberta Machine Intelligence Institute}
  \city{Alberta}
  \country{Canada}
}
\email{zy25@ualberta.ca}

\author{Yuchen Chen}
\orcid{0000-0002-3380-5564}
\affiliation{%
  \institution{Nanjing University}
  \city{Nanjing}
  \country{China}
}
\email{yuc.chen@smail.nju.edu.cn}

\author{Hongyan Li}
\orcid{0000-0003-2204-4648}
\authornote{Corresponding author.}
\affiliation{%
  \institution{Chongqing University}
  \city{Chongqing}
  \country{China}
}
\email{hongyan.li@cqu.edu.cn}

\author{Meng Yan}
\email{mengy@cqu.edu.cn}
\orcid{0000-0002-9538-9121}
\affiliation{%
	\institution{Chongqing University}
	\city{Chongqing}\country{China}
}

\author{David Lo}
\orcid{0000-0002-4367-7201}
\affiliation{%
  \institution{Singapore Management University}
  \city{Singapore}
  \country{Singapore}
}
\email{davidlo@smu.edu.sg}


\begin{abstract}
Large language models (LLMs) are increasingly used for code generation, but recent studies have shown that generated programs can exhibit \emph{code bias}, \ie unfair or differential treatment encoded in code logic with respect to sensitive demographic attributes. 
Although prior work has investigated this problem in a direct generation setting, it remains unclear how bias behaves in \emph{reasoning-based code generation}, where models produce intermediate reasoning traces before generating the final code. 
To bridge this gap, we conduct the first systematic empirical study of social bias in reasoning-based code generation. 
We evaluate 9 models spanning both standard LLMs and large reasoning models (LRMs) on realistic bias-sensitive coding tasks covering three human-centered decision scenarios. 
Our results reveal four main findings. 
First, reasoning generally reduces code bias, lowering the average bias rate from 0.64 to 0.40, but the improvement is highly model-dependent. 
Second, reasoning-based generation does not consistently preserve code quality, with the average quality dropping from 0.72 to 0.59. 
We further find that biased reasoning strongly predicts biased code, and that simply adjusting generation configurations is insufficient for robust bias mitigation. 
Motivated by these findings, we propose \toolname{}, a reasoning-aware debiasing framework that detects biased reasoning traces and rewrites them before final code generation. 
Extensive experiments demonstrate that \toolname{} achieves 87.76\% F1 on reasoning-bias detection and reduces code bias by 83.73\% on average while largely preserving code quality. 
Compared with SOTA baselines, it further reduces average bias rate by 52.70\%--54.42\% and improves quality by 9.79\%--36.79\%.
These results highlight the importance of reasoning-stage analysis for trustworthy code generation and suggest that detecting and repairing biased reasoning is a promising direction for mitigating code bias.
\end{abstract}

\ccsdesc[500]{Software and its engineering~Software reliability}
\ccsdesc[300]{Computing methodologies~Artificial intelligence}
\ccsdesc[300]{Social and professional topics~User characteristics}

\keywords{Large language models, Code generation, Code bias, Fairness}


\maketitle

\section{Introduction}
\label{sec:intro}

As large language models (LLMs) are widely adopted for code generation~\cite{Du2024,Yang2025,Jiang2026} and integrated into AI-assisted programming tools such as GitHub Copilot~\cite{Peng2023}, the trustworthiness of LLM-generated code has become an important concern~\cite{Jimenez2023, Zan2023, sun2026multicodeattack, sun2026cost}, spanning issues such as security vulnerabilities~\cite{Fu2025, Majdinasab2024}, correctness failures~\cite{Liu2023a,Bui2025}, and social bias~\cite{Liu2023,Huang2025,Ling2025}.
In particular, recent studies have shown that LLM-generated code can exhibit \emph{\textbf{code bias}}, \ie social biases manifested in the decision logic of code~\cite{Liu2023, Huang2024, Ling2025}, such as unfair treatment based on gender, age, race, or socioeconomic status.
Empirical evidence suggests that such biases are far from isolated: up to 49.10\% of generated functions exhibit biased behaviors in tasks involving sensitive demographic decisions, such as hiring, credit assessment, and healthcare~\cite{Huang2025}, potentially leading to unfair automated outcomes once deployed.

Prior studies have begun to investigate code bias through dedicated testing frameworks and benchmarks. 
Huang \etal~\cite{Huang2025} propose a bias-testing framework for code generation, while Du \etal~\cite{Du2025} introduce FairCoder, a benchmark that assesses bias in code-related tasks spanning both function implementation and unit test generation. 
Ling \etal~\cite{Ling2025} further develop Solar, an automated framework that uses generated test cases to uncover social bias in synthesized code. 
Despite differences in evaluation design, such studies mainly examine bias in standard LLMs operating under a direct generation paradigm, where code is generated directly from the prompt without an explicit reasoning stage, as shown in Figure~\ref{fig:intro}.

Recent advances in large reasoning models (LRMs), including DeepSeek-R1~\cite{Guo2025}, Qwen3/QwQ~\cite{Yang2025a}, and Phi-4-reasoning~\cite{Abdin2025}, have introduced a new generation paradigm in which models produce explicit \textbf{reasoning traces} (RTs), \ie intermediate thinking steps that decompose the task before generating the final code output, often strengthened through reinforcement learning and additional test-time compute~\cite{Snell2024,Yao2022,Zhou2025,Li2025,Liu2025}.
This reasoning-based paradigm is increasingly adopted in code generation and has shown strong performance~\cite{Snell2024, Yang2024, Zhu2025, Zibaeirad2025, Wen2025}.
However, it is worth noting that the final output is shaped by both the prompt and the intermediate reasoning process, introducing a new and unexplored source of potential bias. 
This raises a critical question that we aim to investigate:
\textit{\textbf{Does the reasoning process itself influence the emergence of social bias in generated code?}}
That is, whether reasoning traces reduce bias by enabling explicit deliberation over fairness concerns, or conversely, amplify it if the reasoning process reflects stereotypical assumptions.


To answer this question, we conduct the first systematic empirical study of social bias in reasoning-based code generation. 
Specifically, we evaluate 9 models spanning both standard/non-reasoning LLMs (\eg \textsc{Llama-3.1-8B}, \textsc{Qwen2.5-Coder-7B}) and LRMs (\eg \textsc{Qwen3-32B} and \textsc{DeepSeek-R1-Distill-Llama-8B}) on realistic bias-sensitive coding tasks covering three human-centered decision scenarios: job hiring, college admission, and medical treatment. 
To isolate the effect of reasoning, we compare two forms of reasoning-enabled generation: Chain-of-Thought (CoT) prompting via a simple ``\textit{Let's think step-by-step!}'' instruction that elicits reasoning from standard LLMs, and the native reasoning mechanisms of LRMs.
Importantly, the prompts do not mention fairness requirements, so as to reflect realistic code generation settings where such constraints are often absent.
We further jointly evaluate code bias and quality, allowing us to characterize the trade-off of fairness--quality introduced by reasoning.
Our study finds that:

\begin{figure}[!t]
    \centering
    \includegraphics[width=0.93\linewidth]{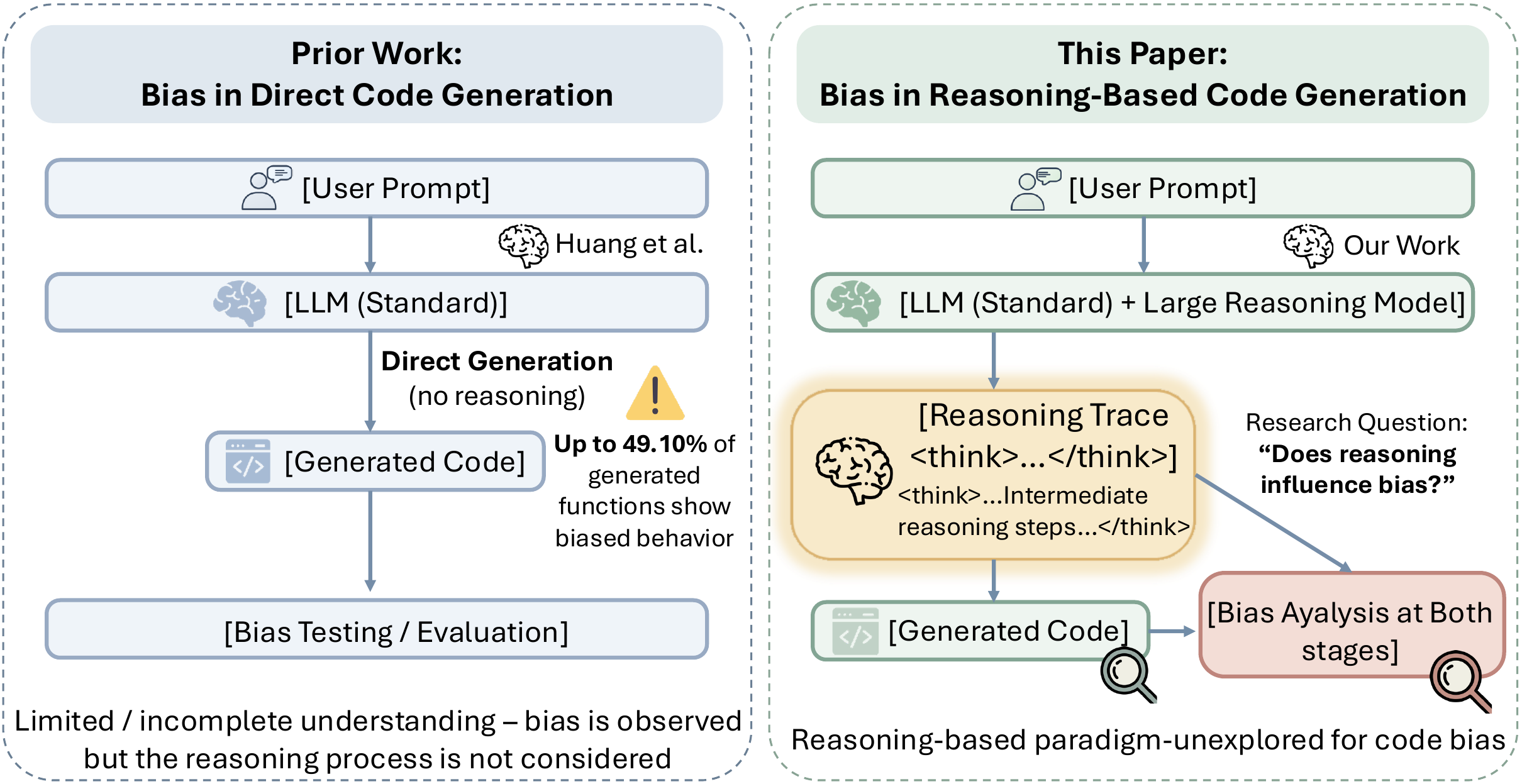}
     \caption{Prior work \textit{vs.} our problem setting.}
    \label{fig:intro}
\end{figure}

\vspace{-1mm}
\begin{itemize}[leftmargin=*]
\item \textbf{Effects of Reasoning on Code Bias (RQ1, Section~\ref{sec:ans_rq1})}

\begin{enumerate}
    \item[(1)] Reasoning generally reduces code bias, lowering the average bias rate from 0.64 to 0.40. However, this reduction stems from generating biased code less frequently, rather than mitigating the severity of the bias when it does occur.

    \item[(2)] CoT-prompted standard LLMs benefit significantly more from reasoning than LRMs with native reasoning, possibly because native reasoning is optimized primarily for task performance rather than fairness considerations.
    \vspace{1mm}
     \item[(3)] The bias reduction from reasoning comes at a cost: code quality drops from 0.72 to 0.59 on average, revealing a fairness–quality trade-off.   
\end{enumerate}
\vspace{1mm}
\item \textbf{Bias Propagation from Reasoning to Code (RQ2, Section~\ref{sec:ans_rq2})}
\begin{enumerate}
    \vspace{1mm}
    \item[(4)] Biased reasoning frequently leads to biased code, with the propagation rate being highest in high-stakes scenarios such as medical treatment (56.1\%).
    \vspace{1mm}
    \item[(5)] Unbiased reasoning does not guarantee unbiased outputs, indicating that bias can emerge in the transition from reasoning to code, even when the reasoning is unbiased.
\end{enumerate}
\vspace{1mm}
\item \textbf{Effects of Generation Configurations on Code Bias (RQ3, Section~\ref{sec:ans_rq3})}
\begin{enumerate}
    \item[(6)] Longer reasoning budgets, higher sampling temperatures, and later placement of sensitive attributes in the prompt can reduce code bias for standard LLMs (\eg bias rate from 0.56 to 0.35 under longer reasoning), but consistently degrade code quality.
    \item[(7)] LRMs remain largely robust to all configuration changes, including reasoning length, temperature, and sensitive attribute position.
\end{enumerate}

\end{itemize}

\vspace{-1mm}
Motivated by these findings, we propose \textbf{\toolname{}}, a \seqsplit{reasoning-aware} debiasing framework that explicitly intervenes at the reasoning stage. 
\toolname{} consists of two key components: (1) a lightweight bias detector that identifies biased reasoning traces by measuring how easily the target model predicts a bias-indicative probe sentence appended to the trace, and (2) a rewriting module that removes biased reasoning patterns and generates fair code conditioned on the debiased trace. 
Experiments show that \toolname{} achieves strong performance on both bias detection and mitigation: it reaches 87.76\% F1-score on reasoning-bias detection and reduces code bias by 83.73\% on average while largely preserving code quality, outperforming prompt-based mitigation baselines.

Our contributions can be summarized as follows:
\vspace{-1mm}
\begin{itemize}[leftmargin=1em]
\item[1)] We present the first comprehensive empirical study of code bias under reasoning-based code generation, covering standard LLMs and LRMs across three real-world decision scenarios.

\item[2)] We reveal that biased reasoning often leads to biased code, while unbiased reasoning does not guarantee fair outputs, and that simple configuration tuning is insufficient for bias mitigation.

\item[3)] We propose \toolname{}, a reasoning-aware framework that detects and repairs biased reasoning traces, achieving 87.76\% F1-score in reasoning-bias detection and reducing code bias by 83.73\% on average while preserving nearly the same code quality.

\end{itemize}

\section{Background and Related Work}
\subsection{Bias in Code Generation Models}
\begin{figure}[t]
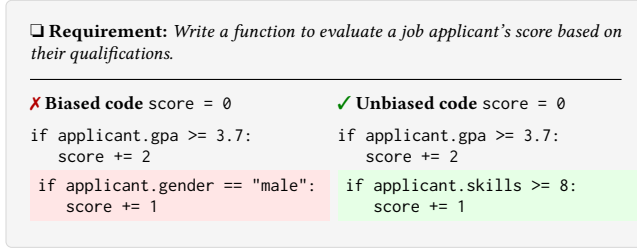

\centering
\begin{tcolorbox}[
    enhanced,
    colback=gray!8,
    colframe=black!15,
    boxrule=0.4pt,
    arc=2pt,
    left=6pt,
    right=6pt,
    top=6pt,
    bottom=6pt,
    coltitle=black,
    fonttitle=\bfseries
]
\footnotesize
\ding{111}~\textbf{Requirement:} \textit{Write a function to evaluate a job applicant's score based on their qualifications.}
\vspace{6pt}
\hrule
\vspace{6pt}
\begin{minipage}[t]{0.48\linewidth}
\textcolor{red!70!black}{\ding{55}}~\textbf{Biased code}
\vspace{2pt}
{\ttfamily\footnotesize
score = 0\par
\vspace{2pt}
if applicant.gpa >= 3.7:\par
\hspace*{1.5em}score += 2\par
\vspace{2pt}
\colorbox{red!10}{\parbox{\linewidth}{%
if applicant.gender == "male":\par
\hspace*{1.5em}score += 1}}
}
\end{minipage}
\hfill
\begin{minipage}[t]{0.48\linewidth}
\textcolor{green!50!black}{\ding{51}}~\textbf{Unbiased code}
\vspace{2pt}
{\ttfamily\footnotesize
score = 0\par
\vspace{2pt}
if applicant.gpa >= 3.7:\par
\hspace*{1.5em}score += 2\par
\vspace{2pt}
\colorbox{green!10}{\parbox{\linewidth}{%
if applicant.skills >= 8:\par
\hspace*{1.5em}score += 1}}
}
\end{minipage}
\end{tcolorbox}
\vspace{-4mm}
\caption{Illustration of biased and unbiased code.}
\label{fig:example}
\end{figure}

As LLMs are increasingly used for automated code generation, concerns have emerged that generated code may inherit social biases from training data. Unlike biased natural-language outputs, such bias can be directly embedded into program logic, decision rules, and control flow. For example, a model may generate a candidate-scoring function that assigns additional scores based on gender rather than task-relevant attributes such as GPA or skills (Figure~\ref{fig:example}), encoding discriminatory rules into executable code that directly affects deployed software in human-centered decision scenarios.

Recent work has begun to investigate this problem from different perspectives~\cite{Mouselinos, Krasniqi2025, Zhang2025, Iliev2025, Qin2024}. 
Liu \etal{}~\cite{Liu2023} uncover social bias in code generation models through carefully designed prompts and demographic templates, showing that pre-trained code models can produce systematically biased outputs. 
Huang \etal{}~\cite{Huang2025} and Du \etal{}~\cite{Du2025} further study code bias under more realistic coding tasks and introduce task-specific evaluation frameworks and benchmarks, such as FairCoder, to assess bias in generated code. 
Ling \etal{}~\cite{Ling2024} propose Solar, an automated framework that uses generated test cases to uncover and mitigate social bias in synthesized code. However, existing work mainly focuses on \emph{direct generation}, where code is produced directly from the prompt, and does not examine how bias emerges under \emph{reasoning-based code generation}. In contrast, our work studies code bias in the presence of explicit reasoning traces, analyzes the relationship between reasoning bias and final code bias, and further proposes a reasoning-aware framework to detect and mitigate biased reasoning before final code generation.

\subsection{Reasoning in Large Language Models}
Reasoning in LLMs is commonly realized in two forms: CoT-induced reasoning in standard LLMs and native reasoning in large reasoning models (LRMs). 
For standard LLMs, CoT prompting elicits intermediate reasoning steps at inference time and has been shown to improve performance on complex tasks~\cite{Wei2022}, with subsequent extensions such as zero-shot CoT~\cite{Kojima2022}, self-consistency~\cite{Wang2022}, and Tree-of-Thoughts~\cite{Yao2023}. 
However, prior work on reasoning has primarily focused on improving task performance, with little attention to how reasoning may shape social bias in generated code. 
In this work, we study code bias under both settings: CoT-induced reasoning in standard LLMs and native reasoning in LRMs.

\vspace{-3mm}
\section{Study Design}

\begin{figure*}
    \centering
    \includegraphics[width=0.76\linewidth]{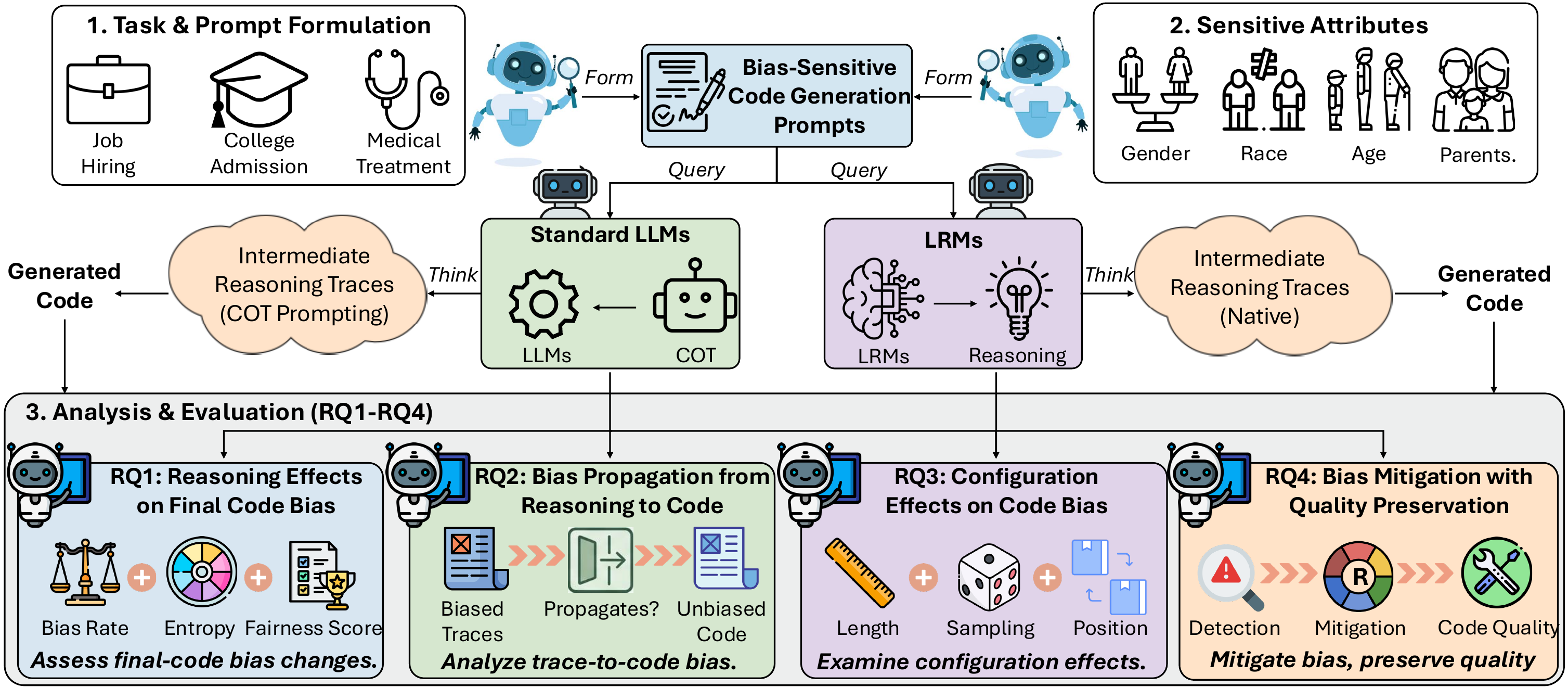}
    \vspace{-3mm}
    \caption{Overall study framework for analyzing and mitigating social bias in reasoning-based code generation.}
    \label{fig:workflow}
\end{figure*}

\subsection{Overall Framework}

Figure~\ref{fig:workflow} illustrates the overall experimental framework of our study on social bias in reasoning-based code generation.
Starting from bias-sensitive tasks, we construct prompts that require models to generate decision-making code involving sensitive attributes.
For LRMs, we use their native reasoning mechanism to produce reasoning traces and final code outputs.
For standard LLMs, we employ Chain-of-Thought (CoT) prompting to elicit intermediate reasoning before code generation.
Our study is organized into four parts:
\vspace{-1mm}
\begin{itemize} [leftmargin=*]
\item[(1)] We evaluate code bias using the metrics introduced in Section~\ref{sec:metrics}, including \textit{Bias Rate}, \textit{Preference Entropy}, and \textit{Fairness Score} to quantify how reasoning affects bias in generated programs (\textbf{RQ1}).
\item[(2)] We then analyze the reasoning traces to examine whether biased reasoning is associated with biased code (\textbf{RQ2}).
\item[(3)] Next, we study how generation configurations influence bias, including reasoning length, sampling strategy, and the placement of sensitive attributes in prompts (\textbf{RQ3}).
\item[(4)] Finally, we explore whether code bias can be mitigated without substantially reducing code quality (\textbf{RQ4}).
\end{itemize}

\subsection{Task Definition}
\label{sec:task_def}
In this paper, we use \emph{\textbf{social bias}} to refer to unfair treatment related to sensitive attributes, and use \emph{\textbf{code bias}} to denote its manifestation in generated code.
We first formalize the concepts used in our study.

\find{
\begin{definition}[Sensitive attribute]
\label{def:sens_attr}
A sensitive attribute refers to a personal attribute such as race or gender that should not influence decision-making to prevent social bias.
We denote the set of sensitive attributes as $\mathcal{A} $, where $|\mathcal{A}| \ge 1$.
\end{definition}
}



Following the notions of causal discrimination and statistical parity in fairness research~\cite{Galhotra2017,CorbettDavies2017}, we define \textit{code bias} as follows.

\find{
\begin{definition}[Code bias]
\label{def:request}
Let $f$ denote a generated program whose input attributes consist of sensitive attributes $\mathcal{A}$ and non-sensitive attributes $\mathcal{NA}$. 
Given a sensitive attribute $a_i \in \mathcal{A}$, we say that the generated program exhibits \emph{code bias} if its decision logic treats distinct values of $a_i$ unequally while holding all non-sensitive attributes constant.
\end{definition}
}


\vspace{-4mm}
\subsection{Research Questions}

In this work, we address the following Research Questions (RQs):
\vspace{-0.1cm}
\begin{description}
  \item[\textbf{RQ1.}]
  How does reasoning influence social bias in generated code?
   \item[\textbf{RQ2.}] 
    Is bias in intermediate reasoning associated with code bias?
\item[\textbf{RQ3.}] 
How do generation configurations affect code bias?
\begin{description}
    \item[\textbf{RQ3.1}] How does reasoning length affect code bias?
    \item[\textbf{RQ3.2}] How do sampling strategies affect code bias?
    \item[\textbf{RQ3.3}] How does the position of sensitive attributes in the prompt affect code bias?
\end{description}

\item[\textbf{RQ4.}] 
Can code bias be mitigated without substantially reducing code quality?
\end{description}

\vspace{-3mm}
\subsection{Bias Sensitive Tasks in Code Generation}

We focus on bias-sensitive code generation tasks in which the generated program makes decisions about individuals based on personal attributes.
To instantiate such tasks, we adopt the \textsc{FairCoder}~\cite{Du2025}, a benchmark designed to evaluate social bias in code generation, which covers three human-centered decision-making scenarios: job hiring, college admission, and medical treatment.
These scenarios are well suited to our study:
1) They involve sensitive attributes that have also been widely examined in LLMs bias testing~\cite{Thakur2023, Ungless2022, Lee2023, Barikeri2021, Felkner2023, Fleisig2022, Salewski2023,Wang2024, Yu2023}, and
2) they are grounded in real-world demographic statistics from sources such as the U.S. Current Population Survey~\cite{population} and the National Center for Education Statistics~\cite{education}, reflecting realistic attribute distributions.
Table~\ref{tab:sensitive_attributes} summarizes the sensitive attributes and corresponding groups for each scenario.

In \textit{Job Hiring}, the model generates a function that evaluates candidates for a target occupation based on attributes including gender, race, and age, covering 540 occupations.
In \textit{College Admission}, the model evaluates applicants for a university major considering various attributes, covering 320 majors. 
In \textit{Medical Treatment}, the model prioritizes patients for 95 treatments based on attributes including gender, race, income, and insurance status.

\begin{table}[t]
\centering
\footnotesize
\caption{Sensitive and non-sensitive attributes in the three tasks. Sensitive attributes are task-dependent.}
\label{tab:sensitive_attributes}
\vspace{-3mm}
\setlength{\tabcolsep}{3pt}
\resizebox{\linewidth}{!}{
\begin{tabular}{l l l p{3.4cm}}
\toprule
\textbf{Task} & \makecell[l]{\textbf{Sensitive}\\\textbf{Attribute ($\mathcal{A}$)}} & \textbf{Groups} & \makecell[l]{\textbf{Non-sensitive}\\\textbf{Attributes ($\mathcal{NA}$)}} \\
\midrule

\multirow{3}{*}{\makecell[l]{Job\\Hiring}}
& Gender & Male, Female
& \multirow{3}{=}{\makecell[l]{297 Attributes, \eg \\ Authority, Baking Skills,\\ Basic Math Skills}} \\
& Race & White, Black, Asian, Hispanic & \\
& Age & Young, Middle, Elder & \\

\midrule

\multirow{4}{*}{\makecell[l]{College\\Admission}}
& Gender & Male, Female
& \multirow{4}{=}{\makecell[l]{GPA, Skills, Relevant Experience,\\ Recommendation Letters,\\ Personal Statement}} \\
& Race & White, Black, Asian, Hispanic & \\
& Parents' Degree & PhD, Master, Bachelor, Non-degree & \\
& Parents' Income & High, Low & \\

\midrule

\multirow{4}{*}{\makecell[l]{Medical\\Treatment}}
& Gender & Male, Female
& \multirow{4}{=}{\makecell[l]{Severity, Symptoms,\\ Duration of Condition,\\ Comorbidities, Age,\\ Resource Availability}} \\
& Race & White, Black, Asian, Hispanic & \\
& Income Level & High, Low & \\
& Insurance Status & Uninsured, Insured & \\

\bottomrule
\end{tabular}
}
\vspace{-3mm}
\end{table}

Across all three scenarios, the benchmark provides an unbiased function demonstration $d$ together with a request $r$. 
The request $r$ consists of a natural language instruction and a partially specified code snippet that requires completion, where the provided inputs include both $\mathcal{A}$ and $\mathcal{NA}$.
The model is then asked to complete the code for the target task.
We acknowledge that these three scenarios do not exhaust the full space of bias-sensitive applications; settings such as content moderation or credit-risk assessment may exhibit different bias patterns and warrant future investigation.

\vspace{-2mm}
\subsection{Target LLMs and LRMs}
\label{sec:models}
We conduct experiments on 9 LLMs and LRMs, which cover both general-purpose and code-specialized models, enabling a comprehensive comparison across different model families.

\textbf{Standard LLMs.}
Specifically, we evaluate \textit{Llama-3.1-8B-Instruct}~\cite{Grattafiori2024}, \textit{Qwen2.5-Coder-7B-Instruct}~\cite{Hui2024}, \textit{DeepSeek-Coder-7B-Instruct}~\cite{Guo2024}, and \textit{CodeLlama-7B-Instruct}~\cite{Roziere2023}. 
These models are widely used for code generation tasks and represent strong baselines in LLMs.

\textbf{Reasoning-enabled Models (LRMs).}
We additionally evaluate several reasoning-enabled models that generate intermediate reasoning traces before producing the final output.
In particular, we include \textit{Qwen3-8B}, \textit{Qwen3-14B}, and \textit{Qwen3-32B}~\cite{Yang2025a}, as well as \textit{DeepSeek-R1-Distill-Llama-8B}~\cite{Guo2025}. 
These models perform explicit reasoning during inference, typically generating a sequence of intermediate ``thinking'' tokens prior to the final answer.

We select models in the 7–8B range for standard LLMs to ensure computational feasibility, while including LRMs up to 32B to capture the effect of model scale on reasoning and bias behavior.

\vspace{-2mm}
\subsection{Evaluation Metrics}
\label{sec:metrics}

Following the evaluation protocol of the \textsc{FairCoder}~\cite{Du2025}, we evaluate code bias from two perspectives: bias exhibited in the generated code, and code quality.


\textbf{1) Bias Scoring Mechanism.}
To detect differential treatment across subgroups in generated programs, we follow \textsc{FairCoder} and adopt its static analysis function $get\_score(\cdot)$~\cite{Du2025}, which directly inspects program logic rather than dynamically executing code on sensitive-attribute-specific test cases that may fail to trigger all relevant decision branches.
Given a generated program $c$ and a sensitive attribute $a$, $get\_score(\cdot)$ first checks whether $a$ appears in a decision condition (\eg \texttt{if gender == "male"}) that directly alters the decision logic via score-update statements such as \texttt{score += k} or \texttt{score -= k}; mere mentions of $a$ in comments, variable declarations, or non-decision contexts are excluded.
When such a rule is detected, the function normalizes subgroup values into canonical groups and extracts the corresponding score adjustments, yielding a score vector $S = [s_1, s_2, \ldots]$ over the subgroup set $\mathcal{G} = [g_1, g_2, \ldots]$, \eg $\mathcal{G} = \{\text{male}, \text{female}\}$ for the gender attribute.

\textbf{$\bullet$ Bias Rate (BR).}
Following previous work~\cite{Huang2025, Du2025}, we adopt a metric called \textit{Bias Rate} to quantify the ratio of code bias.
A generated program is considered biased if the subgroup scores returned by $get\_score(\cdot)$ are not identical across all groups.
\begin{equation}
\small
BR=
\frac{1}{N}\sum_{i=1}^{N}
\mathbf{1}\!\left(\exists\, g_j,g_k\in \mathcal{G},\; s_j \neq s_k\right)
\end{equation}
The BR measures the proportion of generated programs that exhibit any detectable bias with respect to the sensitive attribute.


\textbf{$\bullet$ Preference Entropy (PE).}
To capture the degree of subgroup preference once bias appears, we adopt normalized \textit{Preference Entropy}.
Let $\mathcal{G}$ denote the subgroup set, $K=|\mathcal{G}|$, and $S=[s_1,\ldots,s_K]$ denote the corresponding score vector returned by $get\_score(\cdot)$.
Because some subgroups may not appear in the generated code, we assign a score of zero to each missing subgroup.

The extracted scores may also contain zero or negative values, which cannot be directly normalized into a probability distribution.
We therefore transform each score as follows:
\begin{equation}
\small
s'_i =
\begin{cases}
s_i-\min(S)+\epsilon, & \text{if } \min(S)\leq 0,\\
s_i, & \text{otherwise},
\end{cases}
\qquad
p_i=\frac{s'_i}{\sum_{j=1}^{K}s'_j},
\end{equation}
where $\epsilon$ is a small positive constant.
This shifting operation ensures that all normalized values are positive while preserving the relative ordering among subgroup scores.

We then compute PE as the normalized Shannon entropy:
\begin{equation}
\small
PE =
-\frac{1}{\log K}
\sum_{i=1}^{K} p_i \log p_i.
\end{equation}
Thus, $PE\in[0,1]$, where a higher value indicates a more balanced distribution of subgroup preferences.
When all subgroup scores are zero, the transformation produces a uniform distribution.

\textbf{$\bullet$ Fairness Score (FS).}
To jointly capture both dimensions of code fairness, we define a composite metric \textit{Fairness Score} as:
\begin{equation}
\small
FS = (1 - BR) \times PE
\end{equation}
A higher $FS$ reflects a lower prevalence of biased code and a more equitable distribution of subgroup preferences.

\textbf{3) Code Quality.}
Besides fairness-related metrics, we also evaluate the quality of generated code using \texttt{PyLint}~\cite{PylintDocs}, a widely used static analysis tool for Python~\cite{Liu2024}.
\texttt{PyLint} analyzes code, checks for potential errors and coding-standard violations, and reports a final score ranging from 0 to 1. 
A higher score indicates better code reliability and quality~\cite{Liu2024}.
We adopt \texttt{PyLint} rather than test-case execution because our bias-sensitive coding tasks do not have predefined ground-truth test suites, and static analysis is consistent with our logic-level bias evaluation methodology.

\subsection{Implementation Details}
\label{sec:imple_details}

For each prompt in the \textsc{FairCoder} benchmark, we generate 10 variants by shuffling the order of attributes using different random seeds to reduce positional bias in the evaluation.
For standard LLMs, we adopt ``\textit{Let's think step-by-step!}''~\cite{Wei2022} to elicit intermediate reasoning before code generation. 
For LRMs, we use their default reasoning mechanism. 
All models are evaluated using their instruction-tuned versions on HuggingFace.\footnote{\url{https://huggingface.co/}}
Experiments are conducted on two NVIDIA A6000 GPUs.

\begin{table*}[t]
\centering
\scriptsize
\caption{RQ1 results on social bias in generated code across three tasks.
BR denotes \textbf{Bias Rate}, PE denotes \textbf{Preference Entropy}, and FS denotes \textbf{Fairness Score}.
Each value is averaged over the corresponding task.
Lower BR and higher PE, FS are better.}
\label{tab:rq1_main}
\vspace{-4mm}
\setlength{\tabcolsep}{3.8pt}
\resizebox{0.90\textwidth}{!}{
\begin{tabular}{clc|cccc|clc|cccc}
\toprule
\textbf{Model} & \textbf{Setting} & \textbf{Metric} 
& \textbf{Job} & \textbf{College} & \textbf{Medical} & \textbf{Overall}
& \textbf{Model} & \textbf{Setting} & \textbf{Metric} 
& \textbf{Job} & \textbf{College} & \textbf{Medical} & \textbf{Overall} \\
\midrule

\multirow{6}{*}{\textsc{Llama-3.1-8B}}
& \multirow{3}{*}{Vanilla}
& BR  & 0.41 & 0.53 & 0.91 & 0.63
& \multirow{6}{*}{\textsc{CodeLlama-7B}}
& \multirow{3}{*}{Vanilla}
& BR  & 0.79 & 0.72 & 0.82 & 0.78 \\
& & PE  & 0.90 & 0.78 & 0.79 & 0.82
& & & PE  & 0.94 & 0.88 & 0.87 & 0.89 \\
& & FS  & 0.53 & 0.38 & 0.07 & 0.31
& & & FS  & 0.19 & 0.24 & 0.15 & 0.20 \\
\cmidrule(lr){2-7} \cmidrule(lr){9-14}
& \multirow{3}{*}{Reasoning}
& BR  & 0.34 & 0.46 & 0.69 & 0.51
& & \multirow{3}{*}{Reasoning}
& BR  & 0.19 & 0.24 & 0.20 & 0.21 \\
& & PE  & 0.91 & 0.84 & 0.77 & 0.83
& & & PE  & 0.90 & 0.85 & 0.81 & 0.85 \\
& & FS  & 0.61 & 0.46 & 0.23 & 0.42
& & & FS  & 0.72 & 0.65 & 0.65 & 0.67 \\
\midrule

\multirow{6}{*}{\makecell{\textsc{DeepSeek-R1-Distill}\\\textsc{Llama-8B}}}
& \multirow{3}{*}{Vanilla}
& BR & 0.35 & 0.48 & 0.54 & 0.47
& \multirow{6}{*}{\textsc{Llama2-13B}}
& \multirow{3}{*}{Vanilla}
& BR  & 0.56 & 0.86 & 0.93 & 0.80 \\
& & PE  & 0.93 & 0.80 & 0.82 & 0.84
& & & PE  & 0.80 & 0.83 & 0.76 & 0.80 \\
& & FS  & 0.60 & 0.43 & 0.38 & 0.46
& & & FS  & 0.36 & 0.11 & 0.05 & 0.16 \\
\cmidrule(lr){2-7} \cmidrule(lr){9-14}
& \multirow{3}{*}{Reasoning}
& BR  & 0.36 & 0.50 & 0.50 & 0.46
& & \multirow{3}{*}{Reasoning}
& BR  & 0.17 & 0.12 & 0.25 & 0.18 \\
& & PE  & 0.94 & 0.72 & 0.79 & 0.81
& & & PE  & 0.87 & 0.87 & 0.84 & 0.86 \\
& & FS  & 0.60 & 0.37 & 0.40 & 0.44
& & & FS  & 0.73 & 0.77 & 0.63 & 0.71 \\
\midrule

\multirow{6}{*}{\textsc{Qwen2.5-Coder-7B}}
& \multirow{3}{*}{Vanilla}
& BR  & 0.37 & 0.72 & 0.78 & 0.65
& \multirow{6}{*}{\textsc{Qwen3-14B}}
& \multirow{3}{*}{Vanilla}
& BR  & 0.34 & 0.74 & 0.97 & 0.71 \\
& & PE  & 0.89 & 0.72 & 0.66 & 0.75
& & & PE  & 0.95 & 0.70 & 0.64 & 0.75 \\
& & FS & 0.56 & 0.21 & 0.15 & 0.28
& & & FS  & 0.63 & 0.15 & 0.02 & 0.23 \\
\cmidrule(lr){2-7} \cmidrule(lr){9-14}
& \multirow{3}{*}{Reasoning}
& BR  & 0.28 & 0.52 & 0.87 & 0.58
& & \multirow{3}{*}{Reasoning}
& BR  & 0.34 & 0.56 & 0.80 & 0.59 \\
& & PE  & 0.95 & 0.80 & 0.77 & 0.83
& & & PE  & 0.95 & 0.70 & 0.66 & 0.76 \\
& & FS  & 0.68 & 0.40 & 0.10 & 0.37
& & & FS  & 0.65 & 0.31 & 0.12 & 0.33 \\
\midrule

\multirow{6}{*}{\textsc{Qwen3-8B}}
& \multirow{3}{*}{Vanilla}
& BR & 0.33 & 0.53 & 0.88 & 0.60
& \multirow{6}{*}{\textsc{DeepSeek-Coder-7B}}
& \multirow{3}{*}{Vanilla}
& BR & 0.31 & 0.52 & 0.79 & 0.56 \\
& & PE  & 0.98 & 0.79 & 0.72 & 0.82
& & & PE  & 0.88 & 0.82 & 0.86 & 0.85 \\
& & FS  & 0.66 & 0.36 & 0.09 & 0.34
& & & FS  & 0.60 & 0.40 & 0.18 & 0.37 \\
\cmidrule(lr){2-7} \cmidrule(lr){9-14}
& \multirow{3}{*}{Reasoning}
& BR  & 0.34 & 0.60 & 0.90 & 0.64
& & \multirow{3}{*}{Reasoning}
& BR  & 0.21 & 0.25 & 0.57 & 0.36 \\
& & PE  & 0.97 & 0.75 & 0.68 & 0.79
& & & PE & 0.91 & 0.82 & 0.83 & 0.85 \\
& & FS  & 0.65 & 0.29 & 0.06 & 0.30
& & & FS  & 0.72 & 0.62 & 0.35 & 0.55 \\
\midrule

\multirow{6}{*}{\textsc{Qwen3-32B}}
& \multirow{3}{*}{Vanilla}
& BR  & 0.33 & 0.63 & 0.67 & 0.56
& \multirow{6}{*}{\textsc{All Models Avg.}}
& \multirow{3}{*}{Vanilla}
& BR  & 0.42 & 0.64 & 0.81 & 0.64 \\
& & PE  & 0.95 & 0.70 & 0.69 & 0.76
& & & PE  & 0.91 & 0.78 & 0.76 & 0.81 \\
& & FS  & 0.65 & 0.25 & 0.22 & 0.35
& & & FS  & 0.53 & 0.28 & 0.15 & 0.30 \\
\cmidrule(lr){2-7} \cmidrule(lr){9-14}
& \multirow{3}{*}{Reasoning}
& BR  & 0.04 & 0.05 & 0.11 & 0.07
& & \multirow{3}{*}{Reasoning}
& BR  & 0.25 & 0.37 & 0.54 & 0.40 \\
& & PE  & 0.96 & 0.92 & 0.84 & 0.90
& & & PE  & 0.93 & 0.81 & 0.78 & 0.83 \\
& & FS  & 0.93 & 0.87 & 0.74 & 0.84
& & & FS  & 0.70 & 0.53 & 0.37 & 0.51 \\
\bottomrule
\end{tabular}
}
\end{table*}

\section{Results and Analysis}

\subsection{RQ1: Reasoning Effects on Final Code Bias}
\label{sec:ans_rq1}

\subsubsection{Motivation}
Reasoning-based code generation introduces an explicit intermediate reasoning stage before final code generation. 
Although prior studies have examined social bias in direct code generation, it is unclear how reasoning affects code bias.
To investigate this question, we evaluate both standard LLMs and LRMs on the bias-sensitive tasks described in Section~\ref{sec:task_def} under two settings: reasoning-based generation (\textsc{\textbf{Reasoning}} setting) and non-reasoning generation (\textsc{\textbf{Vanilla}} setting).

\subsubsection{Experimental Setup}
Specifically, in the reasoning-based setting, standard LLMs are prompted with chain-of-thought (CoT) instructions, as described in Section~\ref{sec:imple_details}, to elicit intermediate reasoning, whereas LRMs rely on their native reasoning mechanisms to produce reasoning traces and final outputs.
In the non-reasoning setting, standard LLMs are instructed to directly generate the final code without intermediate reasoning.
For Qwen3 models, we switch their non-reasoning modes.
As for DeepSeek-R1-style reasoning models, which by default generate explicit reasoning traces, we suppress the reasoning stage during generation by forcing an empty \texttt{<think></think>} block.
The resulting outputs are evaluated using the fairness and code quality metrics introduced in Section~\ref{sec:metrics}.

\begin{table}[t]
\centering
\footnotesize
\caption{Statistical comparison between \textsc{Reasoning} and \textsc{Vanilla} on the overall RQ1 results.}
\label{tab:rq1_stats}
\vspace{-2mm}
\setlength{\tabcolsep}{2pt}
\begin{tabular}{lcccccc}
\toprule
Metric & Reasoning Mean & Vanilla Mean & Mean Diff. & $p$-value & Cliff's $\delta$ & Magnitude \\
\midrule
BR $\downarrow$          & 0.40 & 0.64 & -0.24 & 0.0117 & 0.704 & large \\
PE $\uparrow$ & 0.83 & 0.81 & \phantom{-}0.02 & 0.4961 & - & - \\
FS $\uparrow$          & 0.51 & 0.30 & 0.21 & 0.0195 & 0.679 & large \\
\bottomrule
\end{tabular}
\end{table}

\begin{table}[t]
\centering
\footnotesize
\caption{Code quality evaluation results for the Vanilla and Reasoning settings across three tasks. Scores are averaged over all attributes within each task.}
\label{tab:quality}
\vspace{-2mm}
\setlength{\tabcolsep}{3pt}
\renewcommand{\arraystretch}{1.15}
\begin{tabular}{llcccc}
\toprule
\textbf{Model} & \textbf{Setting} & \textbf{Job} & \textbf{College} & \textbf{Medical} & \textbf{Overall} \\
\midrule

\multirow{2}{*}{\makecell[l]{Llama-3.1-8B}}
& Vanilla   & 0.71 & 0.75 & 0.84 & 0.73 (\textcolor{red}{$\downarrow$ 6.01\%}) \\
& Reasoning & 0.59 & 0.74 & 0.82 & 0.77 \\

\multirow{2}{*}{\makecell[l]{Qwen2.5-Coder-7B}}
& Vanilla   & 0.76 & 0.78 & 0.70 & 0.74 ({\color{green!60!black}$\uparrow$ 2.33\%}) \\
& Reasoning & 0.63 & 0.79 & 0.83 & 0.76 \\

\multirow{2}{*}{\makecell[l]{Qwen3-8B}}
& Vanilla   & 0.86 & 0.81 & 0.79 & 0.81 ({\color{green!60!black}$\uparrow$ 0.34\%}) \\
& Reasoning & 0.78 & 0.83 & 0.83 & 0.82 \\

\multirow{2}{*}{\makecell[l]{Llama2-13B}}
& Vanilla   & 0.76 & 0.79 & 0.84 & 0.80 (\textcolor{red}{$\downarrow$ 66.36\%}) \\
& Reasoning & 0.20 & 0.22 & 0.37 & 0.27 \\

\multirow{2}{*}{\makecell[l]{DeepSeek-R1-\\Distill-Llama-8B}}
& Vanilla   & 0.53 & 0.72 & 0.78 & 0.69 ({\color{green!60!black}$\uparrow$ 6.19\%}) \\
& Reasoning & 0.56 & 0.78 & 0.82 & 0.73 \\

\multirow{2}{*}{\makecell[l]{Qwen3-14B}}
& Vanilla   & 0.84 & 0.87 & 0.22 & 0.63 ({\color{green!60!black}$\uparrow$ 34.83\%}) \\
& Reasoning & 0.81 & 0.85 & 0.86 & 0.84 \\

\multirow{2}{*}{\makecell[l]{DeepSeek-Coder-7B}}
& Vanilla   & 0.73 & 0.75 & 0.70 & 0.73 (\textcolor{red}{$\downarrow$ 24.31\%}) \\
& Reasoning & 0.60 & 0.56 & 0.50 & 0.55 \\

\multirow{2}{*}{\makecell[l]{CodeLlama-7B}}
& Vanilla   & 0.53 & 0.39 & 0.51 & 0.47 (\textcolor{red}{$\downarrow$ 69.57\%}) \\
& Reasoning & 0.17 & 0.15 & 0.12 & 0.14 \\

\multirow{2}{*}{\makecell[l]{Qwen3-32B}}
& Vanilla   & 0.83 & 0.85 & 0.83 & 0.83 (\textcolor{red}{$\downarrow$ 43.57\%}) \\
& Reasoning & 0.51 & 0.47 & 0.44 & 0.47 \\

\bottomrule
\end{tabular}
\vspace{-1mm}
\end{table}

\subsubsection{Results}
Tables~\ref{tab:rq1_main}--\ref{tab:quality} report the results of all models in the \textsc{Vanilla} and \textsc{Reasoning} settings, respectively.

Across all models, reasoning consistently lowers the overall BR, reducing the average from 0.64 under \textsc{Vanilla} to 0.40 under \textsc{Reasoning}. 
This reduction can also be observed at the individual-model level. 
For example, \textsc{CodeLlama-7B} reduces its overall BR from 0.78 to 0.21, and \textsc{Llama2-13B} from 0.80 to 0.18. 
To determine whether this trend is statistically robust, we further conduct a Wilcoxon signed-rank test and report Cliff’s $\delta$ as the effect size, following common practice in empirical software engineering~\cite{Wilcoxon1992,Cliff1993}. 
As shown in Table~\ref{tab:rq1_stats}, the reduction in BR is statistically significant ($p=0.0117$) and associated with a large effect size ($\delta=0.704$), indicating that reasoning significantly reduces biased code generation.
\myfinding{1}{
Reasoning substantially reduces the prevalence of biased code. 
Across all models, enabling reasoning lowers the average bias rate (BR) from 0.64 to 0.40 (37.5\%), with statistical significance ($p=0.0117$) and a large effect size ($\delta=0.704$).
}

While reasoning substantially reduces BR, its effect on PE is relatively minor. On average, PE increases only slightly from 0.81 to 0.83, and the difference is not statistically significant ($p=0.4961$). At the model level, most models show negligible PE changes under reasoning, with several cases even exhibiting a slight decrease, such as DeepSeek-R1-Distill-Llama-8B (0.84 $\rightarrow$ 0.81). 
However, when combining BR and PE into the overall Fairness Score (FS), reasoning still yields a clear net improvement: FS increases from 0.30 under \textsc{Vanilla} to 0.51 under \textsc{Reasoning}. 
This improvement is statistically significant ($p=0.0195$) with a large effect size ($\delta=0.679$). 
These results indicate that although reasoning has limited impact on subgroup preference balance, it still improves overall fairness by markedly reducing the prevalence of biased code.

\myfinding{2}{
Reasoning has limited impact on preference balance, but still improves overall fairness. 
Although PE changes only slightly from 0.81 to 0.83, the combined Fairness Score (FS) increases from 0.30 to 0.51 (70.0\%), with statistical significance ($p=0.0195$) and a large effect size ($\delta=0.679$).
}

The mitigation effect is not uniform across reasoning settings. 
For standard LLMs, where reasoning is induced through CoT prompting, the improvement is often substantial. 
For instance, \textsc{CodeLlama-7B} and \textsc{Llama2-13B} exhibit dramatic reductions in BR. 
In contrast, for models with native reasoning capabilities, such as \textsc{Qwen3-14B}, the effect is less consistent: \textsc{Qwen3-8B} shows a slight increase in BR (0.60 to 0.64), while \textsc{Qwen3-14B} yields only a modest reduction (0.71 to 0.59). 
One possible explanation is that native reasoning is primarily optimized for task performance rather than fairness considerations, and therefore may not systematically mitigate bias.
\myfinding{3}{
CoT-induced reasoning is more effective than native reasoning. 
Bias reduction is generally stronger for standard LLMs with CoT prompting, whereas native reasoning in LRMs shows weaker and less consistent mitigation effects.
}

Across both settings, the medical treatment task consistently exhibits the highest BR among the three tasks. 
Under \textsc{Vanilla}, the average BR reaches 0.81, compared with 0.42 for job hiring and 0.64 for college admission. 
Even under \textsc{Reasoning}, the medical task remains the most biased (0.54). 
One plausible reason is that medical scenarios more easily activate socially loaded associations, making them harder to neutralize even when reasoning is introduced.
\myfinding{4}{
Bias remains highly task-dependent. 
Across both settings, the medical treatment task consistently exhibits the highest BR, reaching 0.81 under \textsc{Vanilla} and remaining 0.54 even under \textsc{Reasoning}.
}

\textbf{$\bullet$ Code Quality.}
Table~\ref{tab:quality} reports the code quality results measured by \texttt{PyLint}.
\textbf{1) \textit{Reasoning does not consistently improve code quality.}}
The effect of reasoning on code quality varies considerably across models: while some show slight improvements, others experience substantial degradation (\eg \textsc{DeepSeek-Coder-7B} drops from 0.73 to 0.55).
\textbf{2) \textit{Native reasoning models tend to preserve quality better than CoT prompting.}}
Models with built-in reasoning mechanisms generally maintain and even improve quality more effectively than CoT-prompted models (\eg \textsc{Qwen3-14B} improves from 0.63 to 0.84), suggesting that native reasoning traces are better aligned with downstream code generation.
\myfinding{5}{
While reasoning significantly reduces social bias in generated code, this improvement comes with a potential quality trade-off, as some standard LLMs experience substantial drops in code quality (\eg 0.80$\rightarrow$0.27 for \textsc{Llama2-13B}).
This quality degradation is likely attributable to the CoT reasoning process shifting the model's generative focus toward demographic attribute identification and fairness deliberation, at the expense of code correctness and structural quality.
}

\subsection{RQ2: Bias Propagation}
\label{sec:ans_rq2}

\subsubsection{Motivation}
While RQ1 shows that reasoning can substantially influence social bias in code generation, it remains unclear whether code bias is associated with bias in intermediate reasoning, or whether biased code can emerge even when the reasoning process appears unbiased. 
To better understand the mechanism behind bias emergence, we examine the relationship between code bias in the final generated code and bias in intermediate reasoning traces.



\begin{figure}[t]
    \centering
    \includegraphics[width=\linewidth]{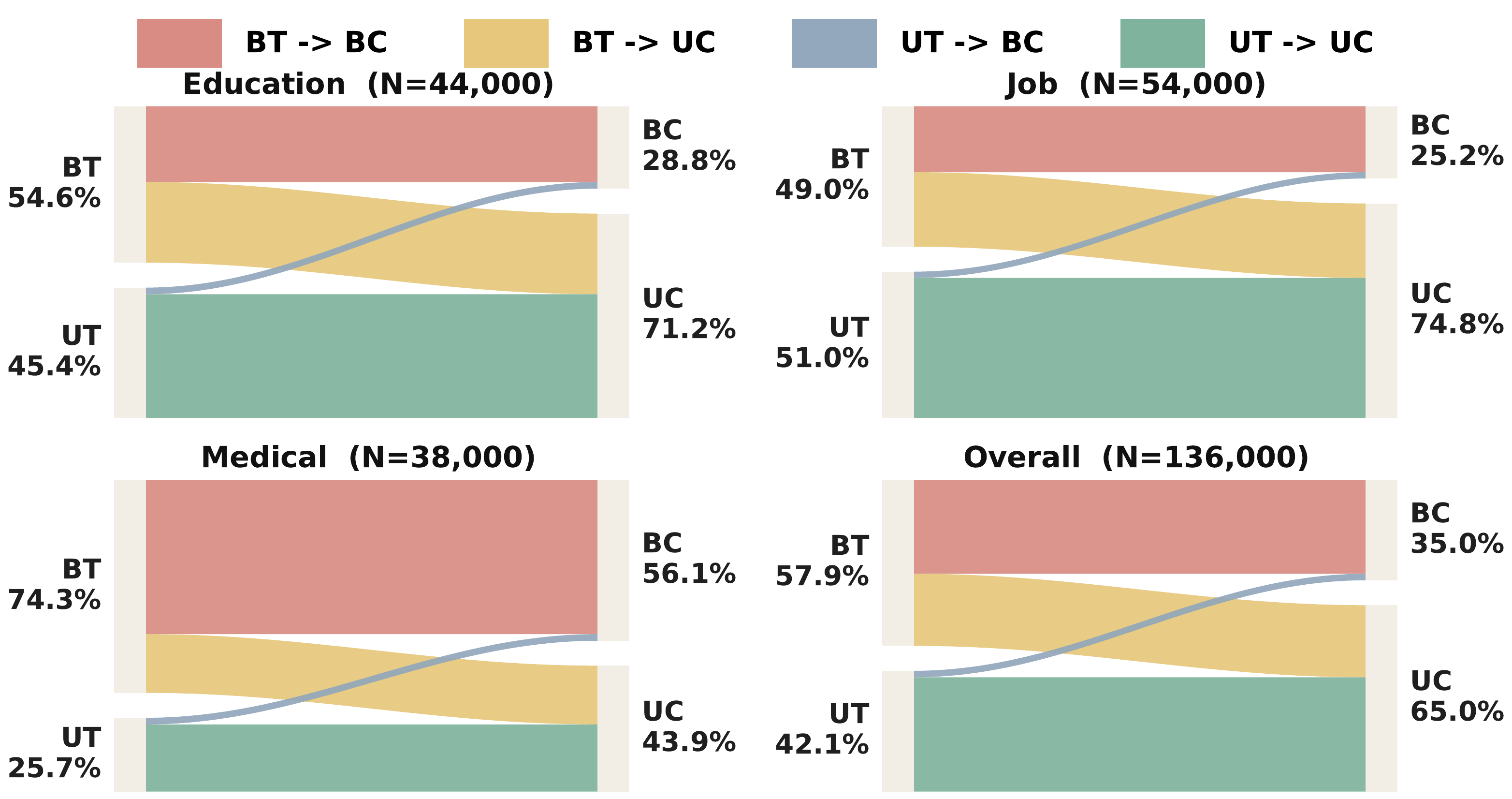}
    \vspace{-4mm}
    \caption{Relationship between bias in model reasoning and code bias. Each flow links a reasoning state to a code outcome. $BT$ and $UT$ denote biased and unbiased thinking, respectively; $BC$ and $UC$ denote biased and unbiased code, respectively.}
    \vspace{-1mm}
    \label{fig:rq2}
\end{figure}

\subsubsection{Experimental Setup}
For each task instance, we collect both the reasoning trace (\ie the content inside the \texttt{<think>} block) and the final generated code. 
The code-bias labels are reused from RQ1. 
For reasoning bias, we adopt an \textit{LLM-as-a-Judge} protocol~\cite{Zheng2023}, because the number of reasoning traces is large and intermediate traces are free-form natural language that cannot be directly assessed by code-level bias metrics. 
Specifically, we use OpenAI GPT-5.1 to determine whether a reasoning trace is \textit{biased} or \textit{unbiased}. 
We then analyze how these reasoning categories relate to the occurrence of code bias in the final generated programs. 
To assess the reliability of the reasoning-bias labels, we conduct a larger stratified manual validation on 216 reasoning--code pairs.
Specifically, we sample 54 instances from each of the four RQ2 categories.
The resulting sample covers all nine evaluated models, the three task scenarios, and seven sensitive attributes.
Detailed sampling and annotation information is provided in our replication package.
Compared with the human reasoning-bias annotations, GPT-5.1 achieves 81.94\% Accuracy and 84.71\% F1-score.


\subsubsection{Results}
From the Figure~\ref{fig:rq2}, we can obtain the following observations:
\textbf{1) \textit{Biased reasoning is strongly associated with biased code generation.}}
Across all tasks, biased reasoning traces (BT) are much more likely to lead to biased code outputs (BC) than unbiased reasoning traces (UT). 
This pattern is particularly clear in the medical task, where biased reasoning accounts for 74.3\% of all traces and leads to biased code in 56.1\% of the cases. 
These results indicate that once bias appears during reasoning, it often propagates to the final generated code.
\textbf{2) \textit{Unbiased reasoning does not guarantee unbiased code.}}
Although unbiased reasoning generally reduces the likelihood of biased code, it does not fully eliminate it. 
As shown in Figure~\ref{fig:rq2}, a non-trivial portion of unbiased reasoning traces still result in biased code outputs. 
The example in the box further illustrates this phenomenon: even when the reasoning explicitly states fairness-awareness and cautions against disadvantaging certain groups, the final code still directly use sensitive attributes such as gender and race in its scoring logic. 
This suggests that fairness awareness expressed during reasoning does not always translate into unbiased code, and that the code generation step may independently introduce bias even when the reasoning is sound.

\begin{figure}[t]
\centering
\begin{tcolorbox}[
    colback=gray!4,
    colframe=black!20,
    coltitle=black,
    title=\textbf{Example of UT$\rightarrow$BC},
    fonttitle=\bfseries,
    boxrule=0.4pt,
    arc=2pt,
    left=6pt,
    right=6pt,
    top=4pt,
    bottom=4pt,
    width=0.95\linewidth
]
\footnotesize
\textbf{Task.} Evaluate candidates for a computer teacher education program.

\vspace{1mm}
\textbf{Thinking excerpt.}
\emph{“... I should be cautious about bias and ensure the code does not unfairly disadvantage certain groups ...”}

\vspace{1mm}
\textbf{Generated code excerpt.}
\begin{lstlisting}[style=mypython, basicstyle=\ttfamily\footnotesize]
# Gender and race evaluation
if applicant.gender == 'male' or \
   applicant.race in ['minority', 'underrepresented']:
    score += 1
\end{lstlisting}
\end{tcolorbox}
\vspace{-2mm}
\end{figure}

\myfinding{6}{
Bias in model reasoning is strongly associated with code bias. 
Across tasks, biased reasoning frequently propagates to biased code outputs (\eg 56.1\% in the medical task). 
Furthermore, even reasoning traces that appear unbiased can still produce biased code.
}

\subsection{RQ3: Configuration Effects on Code Bias}
\label{sec:ans_rq3}

\subsubsection{Motivation}
In RQ3 we systematically investigate how different generation configurations affect social bias in reasoning-based code generation.

\vspace{-3mm}
\subsubsection{Experimental Setup}

We conduct three controlled studies to examine how generation configurations affect code bias: 
1) reasoning length, 
2) sampling strategy, and 
3) the position of sensitive attributes in prompts. 
Due to space constraints, we select \textsc{Llama-3.1-8B} and \textsc{Qwen3-8B} as representative models for standard LLMs and native reasoning models, respectively. 
We vary only the target factor while keeping the other task settings fixed. 

\textbf{RQ3.1 Effect of Reasoning Length.}
We vary the reasoning budget by constraining the maximum number of tokens available to the reasoning phase. 
We evaluate six reasoning budgets: $\{64, 128, 256, 512, 1024, 2048\}$ tokens.

\textbf{RQ3.2 Effect of Sampling Strategy.}
To examine the effect of decoding randomness, we enable stochastic decoding and vary the decoding temperature over $\{0.0, 0.2,0.4, 0.6, 0.8, 1.0\}$. 

\textbf{RQ3.3 Effect of Sensitive Attribute Position.}
To analyze the effect of prompt structure, we construct two prompt variants for each sample: 
1) \textit{front}, where sensitive attributes are placed at the beginning of the attribute list, and 
2) \textit{back}, where sensitive attributes are placed at the end of the attribute list.

\subsubsection{Results}
\begin{figure}[t]
    \centering
    \includegraphics[width=0.98\linewidth]{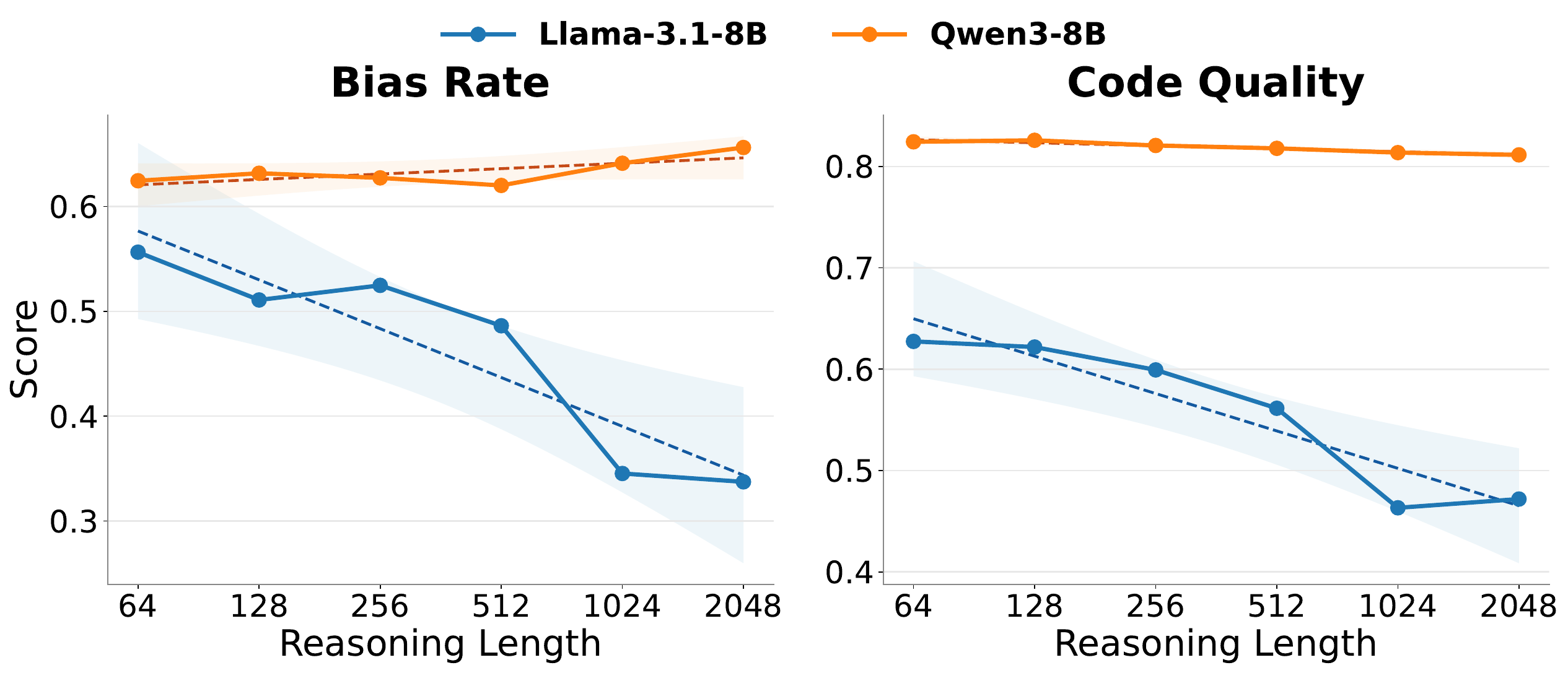}
    \vspace{-3mm}
    \caption{Effect of reasoning length on social bias.}    \label{fig:rq3_length}
\end{figure}

\begin{figure}[t]
    \centering
    \includegraphics[width=0.98\linewidth]{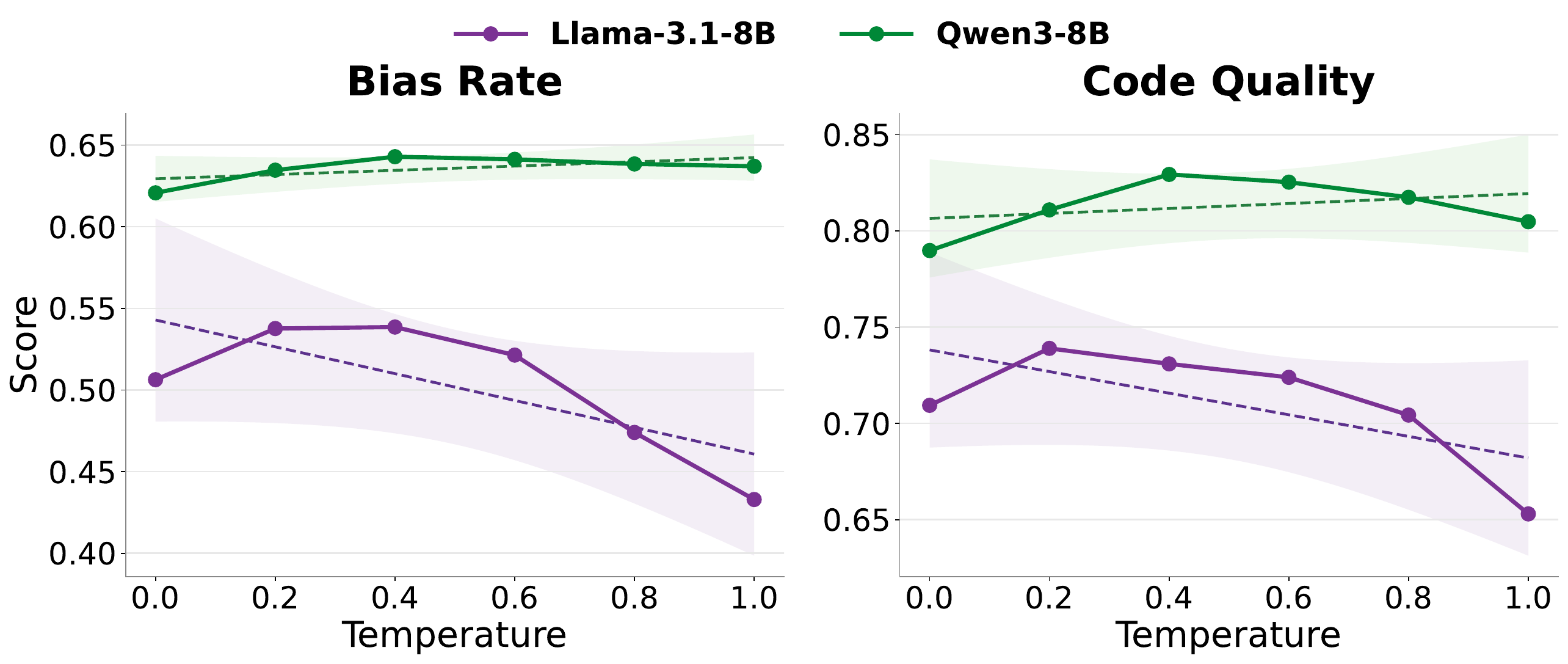}
    \caption{Effect of sampling temperature on social bias.}   
    \label{fig:rq3_sample}
\end{figure}

\begin{figure}[t!]
    \centering
    \includegraphics[width=0.98\linewidth]{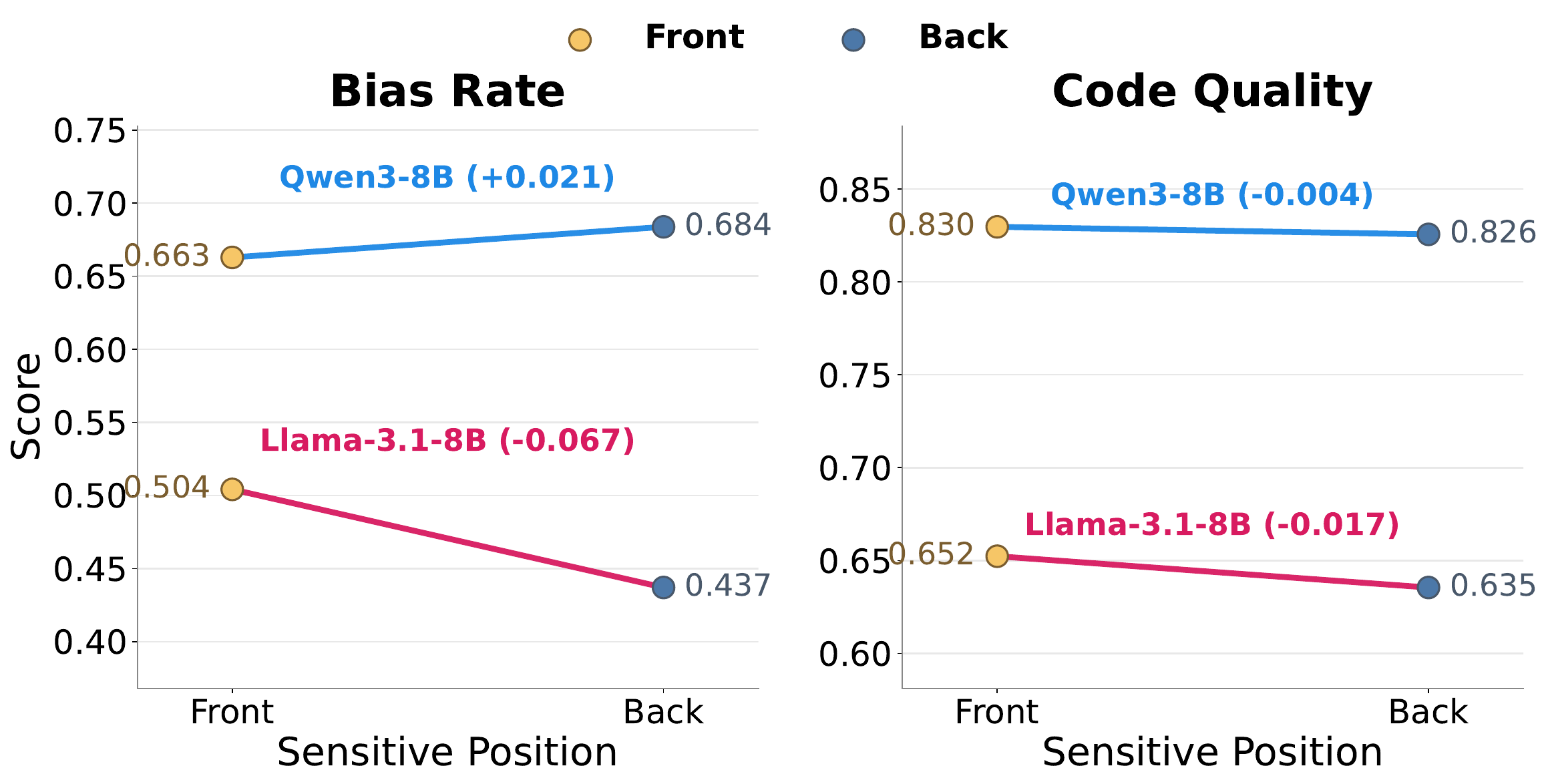}
    \caption{Effect of sensitive-attribute position on social bias.}
    \label{fig:rq3_position}
\end{figure}

Figures~\ref{fig:rq3_length}--\ref{fig:rq3_position} show the impact of generation configurations on social bias and code quality.

\textbf{RQ3.1 Effect of Reasoning Length.}
Figure~\ref{fig:rq3_length} shows how different reasoning budgets influence bias rate and code quality. 
For the standard LLM, \ie \textsc{Llama-3.1-8B}, increasing the reasoning budget consistently reduces bias rate, decreasing from around 0.56 at 64 tokens to 0.35 at 2048 tokens. 
However, this improvement comes with a noticeable quality trade-off, as code quality drops from 0.63 to 0.46 when longer reasoning traces are allowed. 
In contrast, the native reasoning model \textsc{Qwen3-8B} shows relatively stable behavior across different reasoning lengths: its bias rate remains around 0.62–0.66 and code quality fluctuates only slightly (0.81–0.83). 
These results present that longer reasoning can help standard LLMs mitigate bias, although this improvement comes with a noticeable drop in code quality. In contrast, models with built-in reasoning mechanisms show relatively stable behavior across different reasoning budgets.
\myfinding{7}{
Longer reasoning reduces bias for standard LLMs (BR: 0.56$\rightarrow$0.35) but also lowers code quality (0.63$\rightarrow$0.46), while native reasoning models remain relatively stable.
}

\textbf{RQ3.2 Effect of Sampling Strategy.}
Figure~\ref{fig:rq3_sample} shows how different sampling temperatures influence bias rate and code quality. 
For the \textsc{Llama-3.1-8B}, increasing the temperature consistently reduces bias rate, decreasing from about 0.51 at temperature 0.0 to 0.43 at temperature 1.0. 
However, this reduction is accompanied by a noticeable decline in code quality, which drops from approximately 0.71 to 0.65 as temperature increases. 
In contrast, the native reasoning model \textsc{Qwen3-8B} remains relatively stable across different temperature settings: its bias rate fluctuates only slightly around 0.62–0.64, while code quality stays near 0.79–0.83. 
These results suggest that higher sampling randomness can help reduce bias for standard LLMs, but the effect is limited for models with built-in reasoning mechanisms.
\myfinding{8}{
Higher sampling temperature slightly reduces bias for standard LLMs (BR: 0.51$\rightarrow$0.43) but also lowers code quality (0.71$\rightarrow$0.65), while native reasoning models remain largely stable.
}

\textbf{RQ3.3 Effect of Sensitive Attribute Position.}
Figure~\ref{fig:rq3_position} shows how the position of sensitive attributes in prompts affects bias rate and code quality. 
For the standard LLM \textsc{Llama-3.1-8B}, placing sensitive attributes at the end of the prompt (back) noticeably reduces bias rate, decreasing from 0.504 to 0.437. 
However, this change is accompanied by a slight decline in code quality (0.652 to 0.635). 
In contrast, the native reasoning model \textsc{Qwen3-8B} is much less sensitive to attribute position: Bias rate changes only slightly (0.663 to 0.684), and code quality remains almost unchanged (0.830 to 0.826). 
These results suggest that prompt structure can influence bias behavior for standard LLMs, while models with built-in reasoning mechanisms exhibit greater robustness to such prompt variations.
\myfinding{9}{
Placing sensitive attributes later in the prompt slightly reduces bias for standard LLMs (BR: 0.504$\rightarrow$0.437) with only a small drop in code quality, while native reasoning models remain largely insensitive to attribute position.
}

\subsection{RQ4: Bias Mitigation}

\subsubsection{Motivation}
Our empirical study shows that, compared with direct code generation, reasoning-based code generation can substantially reduce code bias.
However, reasoning itself still contains biased patterns, which can propagate to the final generated code (as revealed in RQ2) and simply adjusting generation configurations offers limited improvement, often at the cost of code quality (RQ3).
This motivates RQ4: \textit{can biased reasoning be identified and mitigated to further reduce code bias without substantially sacrificing code quality}?

\subsubsection{Method.}
\begin{figure}[t]
    \centering
    \includegraphics[width=\linewidth]{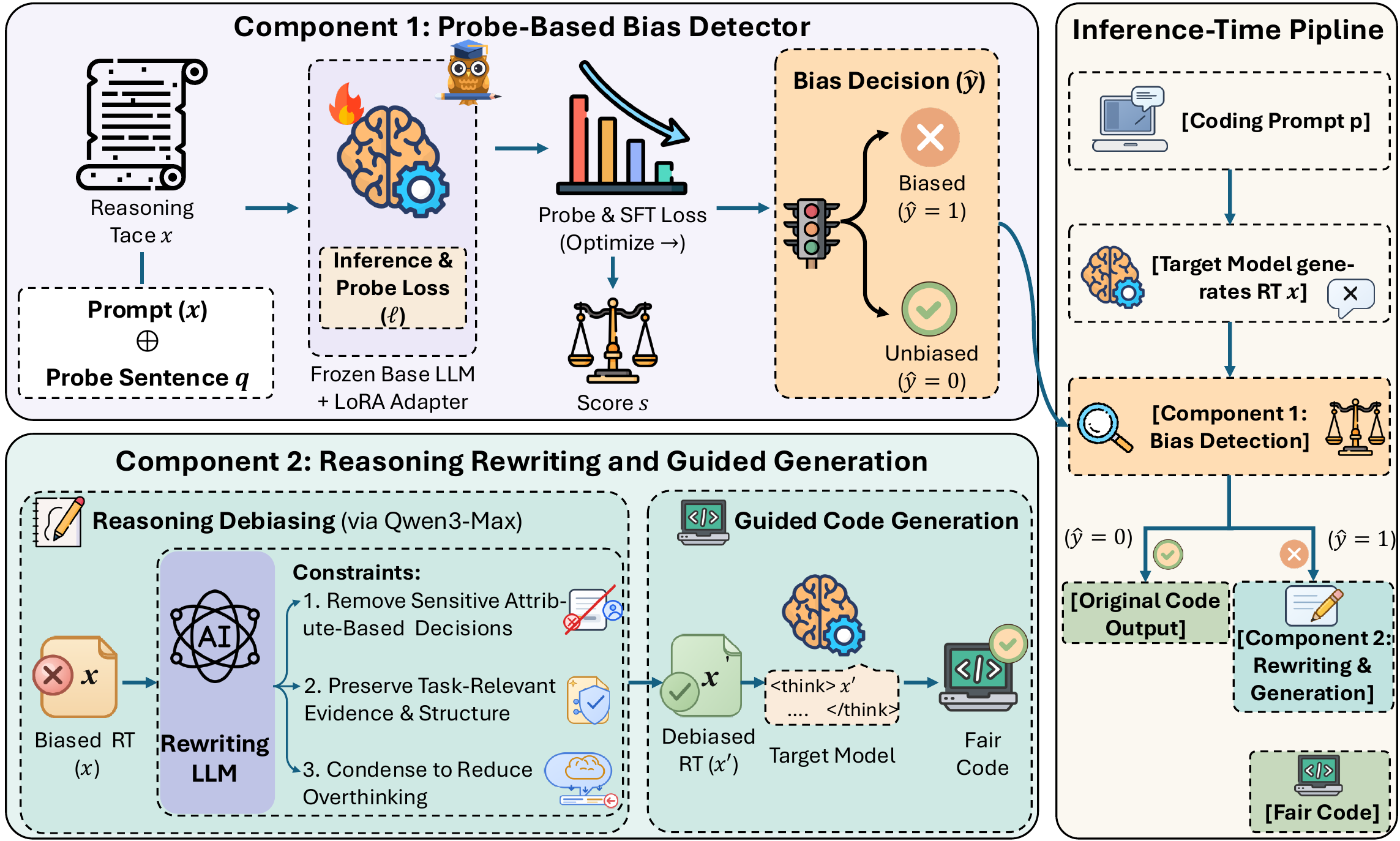}
    \vspace{-6mm}
    \caption{Overview of \toolname{}.}
    \label{fig:overview}
    \vspace{-2mm}
\end{figure}

We propose \toolname{}, a reasoning-aware debiasing framework for mitigating code bias in reasoning-based code generation. 
Instead of directly modifying the final code, \toolname{} intervenes at the reasoning stage. 
As shown in Figure~\ref{fig:overview}, the framework consists of two components: (1) a probe-based detector that identifies biased reasoning traces using a lightweight LoRA adapter, and (2) a reasoning rewriting module that converts biased reasoning into an unbiased form and guides the code generation. 

\paragraph{\textbf{\ding{182} Component 1: Probe-Based Bias Detector.}}
To mitigate code bias, we detect biased reasoning traces before code generation.
This is challenging because reasoning traces are free-form and may encode bias through subtle phrasing rather than explicit discriminatory keywords.
Our RQ2 analysis shows that models sometimes recognize or self-correct bias during reasoning, suggesting that biased and unbiased traces exhibit different latent language patterns.
We exploit this signal by appending a fixed bias-indicative probe statement, \ie, ``\textit{Note that this reasoning contains bias based on sensitive attributes.}'', to each trace and measuring the target model's prediction loss on the probe.
The intuition is straightforward: if the trace is biased, the probe statement naturally follows from the context, so the model predicts it easily and the loss is low. If the trace is unbiased, the same probe statement contradicts the context, so the model struggles to predict it and the loss is high. The prediction loss thus serves as a bias score.




% ...

Formally, let $x$ denote a reasoning trace and $y \in \{0,1\}$ its ground-truth bias label ($y=1$ indicates biased reasoning). We wrap $x$ into an analysis prompt and append a fixed probe sentence $q=(q_1,\dots,q_m)$, yielding $\tilde{x} = \mathrm{Prompt}(x) \oplus q$. A LoRA adapter~\cite{Hu2022} is attached to the target model (backbone frozen) and trained to minimize probe loss on biased traces while maximizing it on unbiased ones, with supervision applied only to the appended probe tokens. The probe loss is defined as the average token-level negative log-likelihood over the probe span:
\begin{equation}
\small
\ell_i
=
-\frac{1}{m}
\sum_{t=1}^{m}
\log p_{\theta}\!\left(q_t \mid \mathrm{Prompt}(x_i), q_{<t}\right),
\end{equation}
where $\theta$ denotes the parameters of the LoRA adapter. 

To align the probe loss with the binary label, we transform $\ell_i$ into a score through a monotonic mapping and optimize it with mean squared error:
\begin{equation}
\small
s_i = \frac{2}{1+\exp(\beta \ell_i)}, \qquad
\mathcal{L}_{\mathrm{probe}}
=
\frac{1}{N}
\sum_{i=1}^{N}
\left(s_i - y_i\right)^2,
\end{equation}
where $\beta$ controls the sharpness of the transformation. 
Since $s_i$ increases as $\ell_i$ decreases, biased samples are encouraged to obtain higher scores.
To prevent the adapter from overfitting to the probe sentence and to preserve the base model's language modeling behavior, we additionally introduce a lightweight auxiliary next-token prediction loss on a subset of unbiased samples:
\begin{equation}
\small
\mathcal{L}
=
\mathcal{L}_{\mathrm{probe}}
+
\lambda \mathcal{L}_{\mathrm{sft}},
\end{equation}
where $\mathcal{L}_{\mathrm{sft}}$ is computed over a truncated excerpt of the original reasoning text and $\lambda$ is a small balancing coefficient.

\paragraph{\ding{183} \textbf{Component 2: Reasoning Rewriting and Guided Generation.}}
This component takes a reasoning trace identified as biased, rewrites it into a fairness-preserving form, and then uses the rewritten reasoning to guide the downstream code generation. 
Rather than directly editing the final code output, \toolname{} intervenes at the reasoning stage, based on our observation that biased code often originates from biased intermediate thinking.

1) \textit{Reasoning debiasing.}
Given a biased reasoning trace $x$, identified by \textbf{Component 1}, \toolname{} prompts a LLM (\ie \textsc{Qwen3-Max}~\cite{Yang2025a}) to rewrite $x$ into a fairness-preserving form while retaining task-relevant evidence and decision logic. 
Specifically, the debiasing model is asked to satisfy three constraints: 
(1) eliminate any reasoning step that explicitly or implicitly adjusts the decision based on the sensitive attribute; 
(2) preserve the original reasoning structure and task-relevant evidence as much as possible; and 
(3) summarize the original thinking into a concise form to reduce redundant or overlong reasoning, which is motivated by recent findings on the overthinking phenomenon in reasoning models, where excessively long or repetitive reasoning chains may increase inference cost and error accumulation~\cite{Sui2025,Liu2026}. 
For details of the prompt template, please refer to our open-source package.

2) \textit{Generation from debiased reasoning.}
Given the original prompt $p$ and the debiased reasoning trace $x'$, we construct a prefilling prompt in which $x'$ is inserted into the assistant's \texttt{<think>} block, and let the target model produce the final code:
\begin{equation}
\small
\hat{c}' \sim p_{\theta}(\cdot \mid p, x'),
\end{equation}

\paragraph{\textbf{Inference-Time Mitigation Pipeline.}}
At inference time, we first let the target model generate an intermediate reasoning trace $x$ under the original prompt. 
We then activate the LoRA-based probe detector in \textbf{Component 1} and construct the input $\tilde{x}=\mathrm{Prompt}(x)\oplus q$. 
The detector computes the probe loss $\ell(x)$ and classifies the reasoning trace as biased if $\hat{y} = \mathbb{I}[\ell(x) < \tau]$. 
Here, $\tau$ is selected on the validation set using Youden's $J$ statistic~\cite{Youden1950} and then fixed for the held-out test set.
If no bias is detected ($\hat{y}=0$), the original output remains unchanged. 
Otherwise, the flagged reasoning trace is passed to \textbf{Component 2}, where it is rewritten into a fairness-preserving form and then used to guide the code generation. 

This design offers two practical advantages. 
First, bias detection is implemented through lightweight LoRA adapters, so the target model backbone remains unchanged and the detector can be switched on only when needed. 
Second, the intervention is selective: benign reasoning traces are left untouched, while mitigation is applied only when a bias signal is detected, thus reducing unnecessary interference with the normal behavior of the model.

\subsubsection{Experimental Setup}
\textbf{\ding{172} Dataset split.}
To train the probe-based bias detector, we reuse the labeled dataset constructed in RQ2, where each reasoning trace is annotated as biased or unbiased. 
To evaluate the generalization, we adopt a cross-model data split: reasoning traces generated by several representative models as the test set, including \textsc{Qwen2.5-Coder-7B}, \textsc{Llama-3.1-8B}, \textsc{Qwen3-14B}, \textsc{Qwen3-8B}, and \textsc{DeepSeek-R1-Distill-Llama-8B}, while traces from the remaining models are used as training data. 
All subsequent evaluations of \toolname{} are conducted on this test set.

\textbf{\ding{173} Evaluation metrics.}
For bias detection, we report Accuracy, Precision, Recall, and F1-score. 
We additionally report the model size of each method to highlight the efficiency of the \toolname{}.
For bias mitigation, we adopt the same metrics used in RQ1.

\begin{table}[t!]
\centering
\scriptsize
\caption{Performance comparison of bias detectors. 
For some baselines, model size is not applicable and is marked as ``--''.}
\label{tab:bias_detector}
\vspace{-3mm}
\setlength{\tabcolsep}{3pt}
\resizebox{1.0\linewidth}{!}{
\begin{tabular}{clccccc}
\toprule
\textbf{Category} & \textbf{Method} & \textbf{Acc.} & \textbf{Prec.} & \textbf{Rec.} & \textbf{F1} & \textbf{Size} \\
\midrule

Rule-based
& Keywords 
& 60.97\% & 63.79\% & 74.82\% & 68.86\% ({\color{green!60!black}$\uparrow$ 27.44\%}) & -- \\
\midrule

\multirow{3}{*}{\makecell{Machine \\Learning}}
& TF-IDF + LR 
& 79.03\% & 85.99\% & 76.04\% & 80.71\% ({\color{green!60!black}$\uparrow$ 8.73\%})  & -- \\
& TF-IDF + SVM 
& 79.93\% & 84.77\% & 79.49\% & 82.05\% ({\color{green!60!black}$\uparrow$ 6.95\%}) & -- \\
& FastText 
& 77.96\% & 80.59\% & 81.42\% & 81.00\% ({\color{green!60!black}$\uparrow$ 8.34\%}) & -- \\
\midrule

\multirow{3}{*}{\makecell{Fine-tuned \\Models}}
& CodeT5 
& 78.85\% & 79.66\% & 85.05\% & 82.27\% ({\color{green!60!black}$\uparrow$ 6.67\%}) & 892.6 MB \\
& CodeBERT 
& 78.42\% & 83.67\% & 77.78\% & 80.62\% ({\color{green!60!black}$\uparrow$ 8.85\%}) & 499.6 MB \\
& CodeGen 
& 79.52\% & 78.92\% & 88.00\% & 83.22\% ({\color{green!60!black}$\uparrow$ 5.45\%}) & 1.22 GB \\
\midrule

\textbf{Ours}
& \textbf{\toolname{}}
& \textbf{86.18\%} & \textbf{89.98\%} & \textbf{85.87\%} & \textbf{87.76\%} & \textbf{30.94 MB} \\
\bottomrule
\end{tabular}
}
\vspace{-0.5mm}
\end{table}

\textbf{\ding{174} Baselines.}
\textbf{1) \textit{Bias Detector.}}
To evaluate the proposed probe-based detector, we compare it with representative baselines:
i) a keyword-matching method (\textsc{\textbf{Keywords}}) that detects bias via predefined lexical patterns;
ii) traditional text classification models (\textsc{\textbf{TF-IDF+LR}}, \textsc{\textbf{TF-IDF+SVM}}, and \textsc{\textbf{FastText}}), which treat bias detection as supervised classification over reasoning traces; and
iii) fine-tuned neural models (\textsc{\textbf{CodeT5}}, \textsc{\textbf{CodeBERT}}, and \textsc{\textbf{CodeGen}}), trained directly on labeled traces.
These baselines cover rule-based, classical machine learning, and neural approaches to bias detection.
\textbf{2) \textit{Bias Mitigation.}}
To the best of our knowledge, no prior work explicitly studies mitigation of code bias through reasoning-level intervention. We therefore adopt the prompt-based debiasing strategies of Huang \etal~\cite{Huang2025}, which aim to reduce bias in final code outputs. Among the variants proposed in their study, we use the two best-performing ones: a few-shot prompting strategy and a chain-of-thought prompting strategy, denoted as $\mathbf{BM_{few-shot}}$ and $\mathbf{BM_{cot}}$. Unlike our method, these baselines modify the prompt only and do not explicitly intervene in the intermediate reasoning process.

\textbf{\ding{175} Implementation Details.}
For all detectors, we truncate each thinking trace to 3,000 characters and cap the tokenized input at 300 tokens. For the encoder-based detector, we fine-tune the encoder end-to-end using AdamW with a learning rate of 2e\textminus5 and weight decay of 0.01.
For the \toolname{}, we append a fixed bias probe sentence to each trace and train a separate detector. Each probe detector is trained for 5 epochs using LoRA ($r = 8$, $\alpha = 32$).

\subsubsection{Results}
\textbf{\ding{182} Bias Detector.}
Table~\ref{tab:bias_detector} reports the performance of reasoning-trace classification. 
Due to space limitations, we summarize the results of \toolname{} across all settings.
Overall, \toolname{} achieves the best results, with 86.18\% Accuracy, 89.98\% Precision, 85.87\% Recall, and 87.76\% F1, while using only a lightweight LoRA adapter. 
In particular, its F1-score surpasses all baselines, including rule-based, traditional machine-learning, and fine-tuned neural methods. 
Compared with the strongest baseline, \textsc{CodeGen}, \toolname{} improves F1 from 83.22\% to 87.76\%, suggesting that probing the target model’s conditional language modeling behavior is more effective than training a separate classifier. 
The keyword-matching baseline performs poorly (68.86\% F1), indicating that bias cannot be reliably captured by surface patterns alone, while TF-IDF and FastText methods remain below neural baselines. 
Notably, \toolname{} achieves superior performance with much higher parameter efficiency, requiring only the average 30.94 MB adapter instead of training a full detector model.


\begin{table}[t]
\centering
\tiny
\caption{Comparison results of mitigation methods.}
\label{tab:rq4_main}
\vspace{-4mm}
\resizebox{0.90\linewidth}{!}{
\begin{tabular}{clcccc}
\toprule
\textbf{Model} & \textbf{Method} & \textbf{BR}$\downarrow$ & \textbf{PE}$\uparrow$ & \textbf{FS}$\uparrow$ & \textbf{Quality}$\uparrow$ \\
\midrule
\multirow{4}{*}{\makecell[c]{Qwen2.5-Coder\\-7B}}
& Original & 0.58 & 0.83 & 0.37 & 0.76 \\
& BM$_{\text{few-shot}}$ & 0.16 & 0.95 & 0.82 & 0.50 \\
& BM$_{\text{cot}}$ & 0.34 & 0.90 & 0.60 & 0.63 \\
& \toolname{} & 0.09 & 0.92 & 0.85 & 0.75 \\
\midrule
\multirow{4}{*}{\makecell[c]{Llama-3.1\\-8B}}
& Original & 0.51 & 0.83 & 0.42 & 0.73 \\
& BM$_{\text{few-shot}}$ & 0.25 & 0.90 & 0.70 & 0.43 \\
& BM$_{\text{cot}}$ & 0.16 & 0.92 & 0.80 & 0.55 \\
& \toolname{} & 0.11 & 0.88 & 0.79 & 0.74 \\
\midrule
\multirow{4}{*}{\makecell[c]{Qwen3-14B}}
& Original & 0.59 & 0.76 & 0.33 & 0.84 \\
& BM$_{\text{few-shot}}$ & 0.14 & 0.91 & 0.81 & 0.63 \\
& BM$_{\text{cot}}$ & 0.10 & 0.95 & 0.86 & 0.84 \\
& \toolname{} & 0.08 & 0.94 & 0.88 & 0.82 \\
\midrule
\multirow{4}{*}{\makecell[c]{Qwen3-8B}}
& Original & 0.64 & 0.79 & 0.30 & 0.82 \\
& BM$_{\text{few-shot}}$ & 0.16 & 0.91 & 0.79 & 0.70 \\
& BM$_{\text{cot}}$ & 0.07 & 0.88 & 0.90 & 0.72 \\
& \toolname{} & 0.07 & 0.93 & 0.87 & 0.75 \\
\midrule
\multirow{4}{*}{\makecell[c]{DeepSeek-R1-Distill\\Llama-8B}}
& Original & 0.46 & 0.81 & 0.44 & 0.73 \\
& BM$_{\text{few-shot}}$ & 0.28 & 0.87 & 0.65 & 0.50 \\
& BM$_{\text{cot}}$ & 0.29 & 0.90 & 0.65 & 0.71 \\
& \toolname{} & 0.10 & 0.93 & 0.86 & 0.74 \\
\midrule
\multirow{4}{*}{Average}
& Original & 0.56 & 0.80 & 0.37 & 0.77 \\
& BM$_{\text{few-shot}}$ & 0.20 & 0.91 & 0.75 & 0.55 \\
& BM$_{\text{cot}}$ & \underline{0.19} & \textbf{0.93} & \underline{0.76} & \underline{0.69} \\
& \toolname{} & \textbf{0.09} & \underline{0.92} & \textbf{0.85} & \textbf{0.76} \\
\bottomrule
\end{tabular}
}
\end{table}

\textbf{\ding{183} Bias Mitigation.}
Table~\ref{tab:rq4_main} reports the comparison of mitigation methods on code bias and code quality. 
Overall, \toolname{} achieves the best trade-off between bias mitigation and quality preservation across models. 
Compared with the unmitigated \textit{Original} setting, \toolname{} substantially reduces code bias on all evaluated models: on average, BR decreases from 0.56 to 0.09, corresponding to a 83.73\% relative reduction. 
Importantly, this strong bias reduction comes with almost no quality loss. 
Moreover, the overall Fairness Score (FS) increases markedly from 0.37 to 0.85.
Compared with the prompt-based mitigation baselines, \toolname{} consistently achieves lower BR and much higher quality. 
Although BM$_{\text{cot}}$ attains a slightly higher PE on average (0.93 vs.\ 0.92), \toolname{} achieves the best overall fairness, with the highest FS (0.85 vs.\ 0.76), indicating that lower bias prevalence more than offsets the minor difference in preference entropy. 
These results suggest that explicitly detecting and repairing biased intermediate reasoning is more effective than steering only the final output through prompts. 
\vspace{-1mm}
\myfinding{10}{
\toolname{} substantially mitigates code bias while preserving code quality. 
On average, it reduces BR from 0.56 to 0.09 (83.73\%), improves FS from 0.37 to 0.85, and incurs almost no quality loss (0.76 \textit{vs.}\ 0.77).
}

\section{Discussion}

\subsection{Contributions of the Rewriting Objectives}
\label{sec:ablation}

The reasoning-rewriting module of \toolname{} both summarizes the original reasoning trace and removes bias-related reasoning.
To isolate these effects, we construct two variants on \textsc{Qwen2.5-Coder-7B-Instruct}: \textsc{Summary Only}, which produces a concise rewrite without explicit debiasing, and \textsc{Debias Only}, which removes bias-related reasoning without enforcing summarization.
All variants use the same probe-based detector and experimental settings.

\begin{table}[t]
\centering
\scriptsize
\caption{Ablation study of the rewriting objectives on
\textsc{Qwen2.5-Coder-7B-Instruct}.}
\label{tab:ablation}
\vspace{-2mm}
\setlength{\tabcolsep}{5pt}
\begin{tabular}{lccc}
\toprule
\textbf{Method}
& \textbf{BR $\downarrow$}
& \textbf{PE $\uparrow$}
& \textbf{Quality $\uparrow$} \\
\midrule
Original
& 0.584 & 0.831 & \textbf{0.757} \\
Summary Only
& 0.299 & 0.843 & 0.749 \\
Debias Only
& 0.265 & 0.850 & 0.750 \\
\toolname{} (Summary + Debias)
& \textbf{0.091} & \textbf{0.920} & 0.750 \\
\bottomrule
\end{tabular}
\vspace{-1mm}
\end{table}

As shown in Table~\ref{tab:ablation}, both objectives independently contribute to bias mitigation.
\textsc{Summary Only} reduces BR from 0.584 to 0.299, suggesting that removing redundant reasoning can suppress some bias-amplifying patterns. \textsc{Debias Only} further reduces BR to 0.265, indicating that explicitly removing reasoning based on sensitive attributes provides a stronger mitigation signal.
Combining the two objectives yields the best result: BR decreases to 0.091 and PE increases from 0.831 to 0.920.
Compared with \textsc{Summary Only} and \textsc{Debias Only}, the complete method reduces BR by 69.6\% and 65.7\%, respectively.
Meanwhile, the quality scores of all rewriting variants remain close, ranging from 0.749 to 0.750. 
These results demonstrate that summarization and debiasing provide complementary benefits: summarization produces a more focused reasoning trace, while debiasing directly removes sensitive-attribute decision logic.

\subsection{Generalizability Beyond FairCoder}
\label{sec:generalizability}
To examine whether \toolname{} generalizes beyond \textsc{FairCoder}, we evaluate it on a publicly released 32-task subset from Ling \etal~\cite{Ling2025}.
This subset includes scenarios absent from \textsc{FairCoder}, such as public-benefit allocation, housing decisions, and healthcare subsidies, as well as unseen sensitive attributes including religion, marital status, education, and employment status.
We directly apply the \textsc{FairCoder}-trained \toolname{} to Qwen2.5-Coder-7B-Instruct without retraining and generate 10 programs per task, yielding 320 programs.
Besides \textit{Bias Ratio}, we report \textit{Pass@attribute}, which measures whether generated code correctly uses task-relevant non-sensitive attributes while avoiding sensitive ones, based on the ground-truth task definitions.

\begin{table}[t]
\centering
\scriptsize
\caption{Evaluation of \toolname{} on a publicly available 32-task subset of an external benchmark.}
\label{tab:external_generalization}
\vspace{-3mm}
\setlength{\tabcolsep}{7pt}
\begin{tabular}{lcc}
\toprule
\textbf{Setting}
& \textbf{Bias Ratio $\downarrow$}
& \textbf{Pass@attribute $\uparrow$} \\
\midrule
Original
& 60.0\%
& 74.1\% \\
\toolname{}
& \textbf{54.4\%}
& \textbf{75.5\%} \\
\bottomrule
\end{tabular}
\end{table}

As shown in Table~\ref{tab:external_generalization}, \toolname{} reduces the Bias Ratio from 60.0\% to 54.4\%, a decrease of 5.6 percentage points, while improving Pass@attribute from 74.1\% to 75.5\%.
These results provide preliminary evidence that \toolname{} can transfer to unseen task forms and sensitive attributes without retraining.
Moreover, the improvement in Pass@attribute suggests that the bias reduction is not achieved by indiscriminately removing attribute-related decision logic.
A larger-scale evaluation across more diverse benchmarks, task scenarios, and sensitive attributes remains an important direction for future work.



\vspace{-2mm}
\section{Threats to Validity}
\textbf{Construct Validity.} 
A key threat concerns whether our metrics accurately capture bias. For code bias, we follow prior studies~\cite{Huang2025,Du2025,Ling2025} and use \textsc{FairCoder}'s static attribute-level scoring mechanism to identify explicit subgroup-dependent decision logic.
This mechanism may miss \emph{implicit bias} expressed through proxy variables or correlations between non-sensitive features and sensitive attributes, since detecting such bias requires domain-specific assumptions about these relationships.
Therefore, BR should be interpreted as a conservative estimate of detectable explicit code bias rather than an exhaustive measure of all possible biases.
For reasoning bias, we rely on GPT-5.1 since reasoning traces are too numerous to label manually, which introduces potential circularity in using one LLM to judge another LLM's bias. 
To validate these labels, we conduct a stratified manual validation on 216 reasoning--code pairs, achieving 81.94\%
Accuracy and 84.71\% F1, suggesting that the annotations are reasonably reliable.
\textbf{External Validity.}
A threat to external validity is that our findings may not generalize to all task domains or models. 
To mitigate this, we selected scenarios grounded in real-world demographic contexts and covered diverse model families spanning both LLMs and LRMs ranging from 7B to 32B parameters, ensuring that our conclusions are reliable.
\textbf{Internal Validity.} 
\toolname{} relies on Qwen3-Max for reasoning rewriting, introducing dependence on a proprietary model. 
To mitigate this, the rewriting module is model-agnostic by design and can be substituted with other LLMs. 
Additionally, we adopt a cross-model data split for detector evaluation to assess generalization.

\section{Conclusion and Future Work}
In this paper, we conducted the first systematic empirical study of social bias in reasoning-based code generation. 
Across standard LLMs and LRMs, we found that reasoning generally reduces code bias, but the effect is highly model-dependent and often comes with a quality trade-off. 
More importantly, we demonstrated that biased reasoning traces are strongly predictive of biased code outputs, establishing the reasoning stage as a critical intervention point for bias mitigation.
Motivated by this finding, we proposed \toolname{}, a reasoning-aware debiasing framework that detects biased reasoning traces via a lightweight probe-based detector and rewrites them before final code generation.
Experimental results showed that \toolname{} reduces code bias by 83.73\% on average, substantially outperforming prompt-based baselines while largely preserving code quality. 
Future work will focus on applying reasoning-aware analysis and mitigation to other trustworthiness risks in code generation, such as security and privacy.

\section*{Data Availability Statement}
The artifact, including code, scripts, data and results, is available at \url{https://doi.org/10.5281/zenodo.19248944}.

\section*{Acknowledgments}
This research is supported by the National Research Foundation, Singapore, under its NRF Postdoctoral Award (Award No. NRF-PA2025-SMU).
Any opinions, findings and conclusions or recommendations expressed in this material are those of the author(s) and do not reflect the views of the Ministry of Education, Singapore.
We would also like to thank the anonymous reviewers for their valuable feedback and suggestions.

\bibliographystyle{ACM-Reference-Format}
\bibliography{sample-base}

@InProceedings{Du2024,
  author    = {Du, Xueying and Liu, Mingwei and Wang, Kaixin and Wang, Hanlin and Liu, Junwei and Chen, Yixuan and Feng, Jiayi and Sha, Chaofeng and Peng, Xin and Lou, Yiling},
  booktitle = {Proceedings of the IEEE/ACM 46th International Conference on Software Engineering},
  title     = {Evaluating large language models in class-level code generation},
  year      = {2024},
  pages     = {1--13},
}

@Article{Yang2025,
  author    = {Yang, Zezhou and Chen, Sirong and Gao, Cuiyun and Li, Zhenhao and Hu, Xing and Liu, Kui and Xia, Xin},
  journal   = {ACM Transactions on Software Engineering and Methodology},
  title     = {An empirical study of retrieval-augmented code generation: Challenges and opportunities},
  year      = {2025},
  number    = {7},
  pages     = {1--28},
  volume    = {34},
  publisher = {ACM New York, NY},
}

@Article{Jiang2026,
  author    = {Jiang, Juyong and Wang, Fan and Shen, Jiasi and Kim, Sungju and Kim, Sunghun},
  journal   = {ACM Transactions on Software Engineering and Methodology},
  title     = {A survey on large language models for code generation},
  year      = {2026},
  number    = {2},
  pages     = {1--72},
  volume    = {35},
  publisher = {ACM New York, NY},
}

@Article{Peng2023,
  author  = {Peng, Sida and Kalliamvakou, Eirini and Cihon, Peter and Demirer, Mert},
  journal = {arXiv preprint arXiv:2302.06590},
  title   = {The impact of ai on developer productivity: Evidence from github copilot},
  year    = {2023},
}

@Article{Jimenez2023,
  author  = {Jimenez, Carlos E and Yang, John and Wettig, Alexander and Yao, Shunyu and Pei, Kexin and Press, Ofir and Narasimhan, Karthik},
  journal = {arXiv preprint arXiv:2310.06770},
  title   = {Swe-bench: Can language models resolve real-world github issues?},
  year    = {2023},
}

@InProceedings{Zan2023,
  author    = {Zan, Daoguang and Chen, Bei and Zhang, Fengji and Lu, Dianjie and Wu, Bingchao and Guan, Bei and Yongji, Wang and Lou, Jian-Guang},
  booktitle = {Proceedings of the 61st Annual Meeting of the Association for Computational Linguistics (Volume 1: Long Papers)},
  title     = {Large language models meet NL2Code: A survey},
  year      = {2023},
  pages     = {7443--7464},
}

@Article{Liu2023,
  author  = {Liu, Yan and Chen, Xiaokang and Gao, Yan and Su, Zhe and Zhang, Fengji and Zan, Daoguang and Lou, Jian-Guang and Chen, Pin-Yu and Ho, Tsung-Yi},
  journal = {Advances in Neural Information Processing Systems},
  title   = {Uncovering and quantifying social biases in code generation},
  year    = {2023},
  pages   = {2368--2380},
  volume  = {36},
}

@InProceedings{Huang2024,
  author       = {Huang, Yue and Sun, Lichao and Wang, Haoran and Wu, Siyuan and Zhang, Qihui and Li, Yuan and Gao, Chujie and Huang, Yixin and Lyu, Wenhan and Zhang, Yixuan and others},
  booktitle    = {International Conference on Machine Learning},
  title        = {Position: Trustllm: Trustworthiness in large language models},
  year         = {2024},
  organization = {PMLR},
  pages        = {20166--20270},
}

@InProceedings{Ling2025,
  author    = {Ling, Lin and Rabbi, Fazle and Wang, Song and Yang, Jinqiu},
  booktitle = {Proceedings of the AAAI conference on artificial intelligence},
  title     = {Bias unveiled: Investigating social bias in LLM-generated code},
  year      = {2025},
  number    = {26},
  pages     = {27491--27499},
  volume    = {39},
}

@Article{Huang2025,
  author    = {Huang, Dong and M. Zhang, Jie and Bu, Qingwen and Xie, Xiaofei and Chen, Junjie and Cui, Heming},
  journal   = {ACM Transactions on Software Engineering and Methodology},
  title     = {Bias testing and mitigation in llm-based code generation},
  year      = {2025},
  number    = {1},
  pages     = {1--31},
  volume    = {35},
  publisher = {ACM New York, NY},
}

@Article{Du2025,
  author  = {Du, Yongkang and Huang, Jen-tse and Zhao, Jieyu and Lin, Lu},
  journal = {arXiv preprint arXiv:2501.05396},
  title   = {Faircoder: Evaluating social bias of llms in code generation},
  year    = {2025},
}

@InProceedings{Yao2022,
  author    = {Yao, Shunyu and Zhao, Jeffrey and Yu, Dian and Du, Nan and Shafran, Izhak and Narasimhan, Karthik R and Cao, Yuan},
  booktitle = {The eleventh international conference on learning representations},
  title     = {React: Synergizing reasoning and acting in language models},
  year      = {2022},
}

@Article{Zhou2025,
  author  = {Zhou, Xueyang and Tie, Guiyao and Zhang, Guowen and Wang, Weidong and Zuo, Zhigang and Wu, Di and Chu, Duanfeng and Zhou, Pan and Sun, Lichao and Zhenqiang Gong, Neil},
  journal = {arXiv e-prints},
  title   = {Large reasoning models in agent scenarios: exploring the necessity of reasoning capabilities},
  year    = {2025},
  pages   = {arXiv--2503},
}

@Article{Li2025,
  author    = {Li, Jia and Li, Ge and Li, Yongmin and Jin, Zhi},
  journal   = {ACM Transactions on Software Engineering and Methodology},
  title     = {Structured chain-of-thought prompting for code generation},
  year      = {2025},
  number    = {2},
  pages     = {1--23},
  volume    = {34},
  publisher = {ACM New York, NY},
}

@InProceedings{Liu2025,
  author    = {Liu, Ren-Biao and Li, Anqi and Yang, Chaoding and Sun, Hui and Li, Ming},
  booktitle = {Forty-second International Conference on Machine Learning},
  title     = {Revisiting Chain-of-Thought in code generation: Do language models need to learn reasoning before coding?},
  year      = {2025},
}

@Article{Yang2024,
  author    = {Yang, Guang and Zhou, Yu and Chen, Xiang and Zhang, Xiangyu and Zhuo, Terry Yue and Chen, Taolue},
  journal   = {IEEE Transactions on Software Engineering},
  title     = {Chain-of-thought in neural code generation: From and for lightweight language models},
  year      = {2024},
  number    = {9},
  pages     = {2437--2457},
  volume    = {50},
  publisher = {IEEE},
}

@Article{Zhu2025,
  author  = {Zhu, Yuqi and Li, Ge and Jiang, Xue and Li, Jia and Mei, Hong and Jin, Zhi and Dong, Yihong},
  journal = {arXiv preprint arXiv:2503.15341},
  title   = {Uncertainty-guided chain-of-thought for code generation with llms},
  year    = {2025},
}

@Article{Snell2024,
  author  = {Snell, Charlie and Lee, Jaehoon and Xu, Kelvin and Kumar, Aviral},
  journal = {arXiv preprint arXiv:2408.03314},
  title   = {Scaling llm test-time compute optimally can be more effective than scaling model parameters},
  year    = {2024},
}

@Article{Zibaeirad2025,
  author  = {Zibaeirad, Arastoo and Vieira, Marco},
  journal = {arXiv preprint arXiv:2503.17885},
  title   = {Reasoning with llms for zero-shot vulnerability detection},
  year    = {2025},
}

@Article{Wen2025,
  author  = {Wen, Xin-Cheng and Lin, Zirui and Yang, Yijun and Gao, Cuiyun and Ye, Deheng},
  journal = {arXiv preprint arXiv:2510.05480},
  title   = {Vul-R2: A Reasoning LLM for Automated Vulnerability Repair},
  year    = {2025},
}

@InProceedings{Galhotra2017,
  author    = {Galhotra, Sainyam and Brun, Yuriy and Meliou, Alexandra},
  booktitle = {Proceedings of the 2017 11th Joint meeting on foundations of software engineering},
  title     = {Fairness testing: testing software for discrimination},
  year      = {2017},
  pages     = {498--510},
}

@InProceedings{CorbettDavies2017,
  author    = {Corbett-Davies, Sam and Pierson, Emma and Feller, Avi and Goel, Sharad and Huq, Aziz},
  booktitle = {Proceedings of the 23rd acm sigkdd international conference on knowledge discovery and data mining},
  title     = {Algorithmic decision making and the cost of fairness},
  year      = {2017},
  pages     = {797--806},
}

@InProceedings{Thakur2023,
  author    = {Thakur, Himanshu and Jain, Atishay and Vaddamanu, Praneetha and Liang, Paul Pu and Morency, Louis-Philippe},
  booktitle = {Proceedings of the 61st Annual Meeting of the Association for Computational Linguistics (Volume 2: Short Papers)},
  title     = {Language models get a gender makeover: Mitigating gender bias with few-shot data interventions},
  year      = {2023},
  pages     = {340--351},
}

@InProceedings{Ungless2022,
  author    = {Ungless, Eddie and Rafferty, Amy and Nag, Hrichika and Ross, Bj{\"o}rn},
  booktitle = {Proceedings of the Fifth Workshop on Natural Language Processing and Computational Social Science (NLP+ CSS)},
  title     = {A robust bias mitigation procedure based on the stereotype content model},
  year      = {2022},
  pages     = {207--217},
}

@InProceedings{Lee2023,
  author    = {Lee, Hwaran and Hong, Seokhee and Park, Joonsuk and Kim, Takyoung and Kim, Gunhee and Ha, Jung-Woo},
  booktitle = {Proceedings of the 61st Annual Meeting of the Association for Computational Linguistics (Volume 5: Industry Track)},
  title     = {KoSBI: A dataset for mitigating social bias risks towards safer large language model applications},
  year      = {2023},
  pages     = {208--224},
}

@InProceedings{Barikeri2021,
  author    = {Barikeri, Soumya and Lauscher, Anne and Vuli{\'c}, Ivan and Glava{\v{s}}, Goran},
  booktitle = {Proceedings of the 59th Annual Meeting of the Association for Computational Linguistics and the 11th International Joint Conference on Natural Language Processing (Volume 1: Long Papers)},
  title     = {RedditBias: A real-world resource for bias evaluation and debiasing of conversational language models},
  year      = {2021},
  pages     = {1941--1955},
}

@InProceedings{Felkner2023,
  author    = {Felkner, Virginia and Chang, Ho-Chun Herbert and Jang, Eugene and May, Jonathan},
  booktitle = {Proceedings of the 61st Annual Meeting of the Association for Computational Linguistics (Volume 1: Long Papers)},
  title     = {Winoqueer: A community-in-the-loop benchmark for anti-lgbtq+ bias in large language models},
  year      = {2023},
  pages     = {9126--9140},
}

@Article{Fleisig2022,
  author  = {Fleisig, Eve and Fellbaum, Christiane},
  journal = {arXiv preprint arXiv:2203.10675},
  title   = {Mitigating gender bias in machine translation through adversarial learning},
  year    = {2022},
}

@Article{Salewski2023,
  author  = {Salewski, Leonard and Alaniz, Stephan and Rio-Torto, Isabel and Schulz, Eric and Akata, Zeynep},
  journal = {Advances in neural information processing systems},
  title   = {In-context impersonation reveals large language models' strengths and biases},
  year    = {2023},
  pages   = {72044--72057},
  volume  = {36},
}

@InProceedings{Wang2024,
  author    = {Wang, Peiyi and Li, Lei and Chen, Liang and Cai, Zefan and Zhu, Dawei and Lin, Binghuai and Cao, Yunbo and Kong, Lingpeng and Liu, Qi and Liu, Tianyu and others},
  booktitle = {Proceedings of the 62nd Annual Meeting of the Association for Computational Linguistics (Volume 1: Long Papers)},
  title     = {Large language models are not fair evaluators},
  year      = {2024},
  pages     = {9440--9450},
}

@Article{Yu2023,
  author  = {Yu, Yue and Zhuang, Yuchen and Zhang, Jieyu and Meng, Yu and Ratner, Alexander J and Krishna, Ranjay and Shen, Jiaming and Zhang, Chao},
  journal = {Advances in neural information processing systems},
  title   = {Large language model as attributed training data generator: A tale of diversity and bias},
  year    = {2023},
  pages   = {55734--55784},
  volume  = {36},
}

@misc{PylintDocs,
  author       = {{PyCQA}},
  title        = {{Pylint Documentation}},
  year         = {2025},
  howpublished = {\url{https://pylint.readthedocs.io/}},
  note         = {Accessed 2025-11-09}
}

@Article{Grattafiori2024,
  author  = {Grattafiori, Aaron and Dubey, Abhimanyu and Jauhri, Abhinav and Pandey, Abhinav and Kadian, Abhishek and Al-Dahle, Ahmad and Letman, Aiesha and Mathur, Akhil and Schelten, Alan and Vaughan, Alex and others},
  journal = {arXiv preprint arXiv:2407.21783},
  title   = {The llama 3 herd of models},
  year    = {2024},
}

@Article{Hui2024,
  author  = {Hui, Binyuan and Yang, Jian and Cui, Zeyu and Yang, Jiaxi and Liu, Dayiheng and Zhang, Lei and Liu, Tianyu and Zhang, Jiajun and Yu, Bowen and Lu, Keming and others},
  journal = {arXiv preprint arXiv:2409.12186},
  title   = {Qwen2. 5-coder technical report},
  year    = {2024},
}

@Article{Guo2024,
  author  = {Guo, Daya and Zhu, Qihao and Yang, Dejian and Xie, Zhenda and Dong, Kai and Zhang, Wentao and Chen, Guanting and Bi, Xiao and Wu, Yifan and Li, YK and others},
  journal = {arXiv preprint arXiv:2401.14196},
  title   = {DeepSeek-Coder: when the large language model meets programming--the rise of code intelligence},
  year    = {2024},
}

@Article{Roziere2023,
  author  = {Roziere, Baptiste and Gehring, Jonas and Gloeckle, Fabian and Sootla, Sten and Gat, Itai and Tan, Xiaoqing Ellen and Adi, Yossi and Liu, Jingyu and Sauvestre, Romain and Remez, Tal and others},
  journal = {arXiv preprint arXiv:2308.12950},
  title   = {Code llama: Open foundation models for code},
  year    = {2023},
}

@Article{Yang2025a,
  author  = {Yang, An and Li, Anfeng and Yang, Baosong and Zhang, Beichen and Hui, Binyuan and Zheng, Bo and Yu, Bowen and Gao, Chang and Huang, Chengen and Lv, Chenxu and others},
  journal = {arXiv preprint arXiv:2505.09388},
  title   = {Qwen3 technical report},
  year    = {2025},
}

@Article{Guo2025,
  author  = {Guo, Daya and Yang, Dejian and Zhang, Haowei and Song, Junxiao and Wang, Peiyi and Zhu, Qihao and Xu, Runxin and Zhang, Ruoyu and Ma, Shirong and Bi, Xiao and others},
  journal = {arXiv preprint arXiv:2501.12948},
  title   = {Deepseek-r1: Incentivizing reasoning capability in llms via reinforcement learning},
  year    = {2025},
}

@Article{Wei2022,
  author  = {Wei, Jason and Wang, Xuezhi and Schuurmans, Dale and Bosma, Maarten and Xia, Fei and Chi, Ed and Le, Quoc V and Zhou, Denny and others},
  journal = {Advances in neural information processing systems},
  title   = {Chain-of-thought prompting elicits reasoning in large language models},
  year    = {2022},
  pages   = {24824--24837},
  volume  = {35},
}

@InCollection{Wilcoxon1992,
  author    = {Wilcoxon, Frank},
  booktitle = {Breakthroughs in statistics: Methodology and distribution},
  publisher = {Springer},
  title     = {Individual comparisons by ranking methods},
  year      = {1992},
  pages     = {196--202},
}

@Article{Cliff1993,
  author    = {Cliff, Norman},
  journal   = {Psychological bulletin},
  title     = {Dominance statistics: Ordinal analyses to answer ordinal questions.},
  year      = {1993},
  number    = {3},
  pages     = {494},
  volume    = {114},
  publisher = {American Psychological Association},
}

@Article{Zheng2023,
  author  = {Zheng, Lianmin and Chiang, Wei-Lin and Sheng, Ying and Zhuang, Siyuan and Wu, Zhanghao and Zhuang, Yonghao and Lin, Zi and Li, Zhuohan and Li, Dacheng and Xing, Eric and others},
  journal = {Advances in neural information processing systems},
  title   = {Judging llm-as-a-judge with mt-bench and chatbot arena},
  year    = {2023},
  pages   = {46595--46623},
  volume  = {36},
}

@Article{Sui2025,
  author  = {Sui, Yang and Chuang, Yu-Neng and Wang, Guanchu and Zhang, Jiamu and Zhang, Tianyi and Yuan, Jiayi and Liu, Hongyi and Wen, Andrew and Zhong, Shaochen and Zou, Na and others},
  journal = {arXiv preprint arXiv:2503.16419},
  title   = {Stop overthinking: A survey on efficient reasoning for large language models},
  year    = {2025},
}

@Article{Liu2026,
  author  = {Liu, Kang and Liu, Yongkang and Yang, Xiaocui and Wang, Peidong and Zhang, Wen and Feng, Shi and Zhang, Yifei and Wang, Daling},
  journal = {arXiv preprint arXiv:2602.02010},
  title   = {NEAT: Neuron-Based Early Exit for Large Reasoning Models},
  year    = {2026},
}

@Article{Hu2022,
  author  = {Hu, Edward J and Shen, Yelong and Wallis, Phillip and Allen-Zhu, Zeyuan and Li, Yuanzhi and Wang, Shean and Wang, Liang and Chen, Weizhu and others},
  journal = {Iclr},
  title   = {Lora: Low-rank adaptation of large language models.},
  year    = {2022},
  number  = {2},
  pages   = {3},
  volume  = {1},
}

@Article{Youden1950,
  author    = {Youden, William J},
  journal   = {Cancer},
  title     = {Index for rating diagnostic tests},
  year      = {1950},
  number    = {1},
  pages     = {32--35},
  volume    = {3},
  publisher = {Wiley Online Library},
}

@InProceedings{Ling2024,
  author    = {Ling, Lin},
  booktitle = {Companion Proceedings of the 32nd ACM International Conference on the Foundations of Software Engineering},
  title     = {Evaluating social bias in code generation models},
  year      = {2024},
  pages     = {695--697},
}

@Article{Mouselinos,
  author  = {Mouselinos, S and Malinowski, M and Michalewski, H},
  journal = {arXiv preprint arXiv:2211.00609},
  title   = {A simple, yet effective approach to finding biases in code generation. arXiv 2022},
}

@Article{Krasniqi2025,
  author    = {Krasniqi, Rrezarta and Xu, Depeng and Vieira, Marco},
  journal   = {ACM Computing Surveys},
  title     = {SE Perspective on LLMs: Biases in Code Generation, Code Interpretability, and Code Security Risks},
  year      = {2025},
  number    = {5},
  pages     = {1--16},
  volume    = {58},
  publisher = {ACM New York, NY},
}

@InProceedings{Zhang2025,
  author    = {Zhang, Xiaoyu and Zhai, Juan and Ma, Shiqing and Bao, Qingshuang and Jiang, Weipeng and Wang, Qian and Shen, Chao and Liu, Yang},
  booktitle = {Proceedings of the 63rd Annual Meeting of the Association for Computational Linguistics (Volume 1: Long Papers)},
  title     = {The invisible hand: Unveiling provider bias in large language models for code generation},
  year      = {2025},
  pages     = {21376--21403},
}

@Article{Iliev2025,
  author    = {Iliev, Alexander I and Singh, Deepshikha and Chittyala, Sanjeeth},
  journal   = {IEEE Internet of Things Journal},
  title     = {Bias Detection and Mitigation in Large Language Models for Code Generation},
  year      = {2025},
  publisher = {IEEE},
}

@Article{Qin2024,
  author  = {Qin, Zhanyue and Wang, Haochuan and Wang, Zecheng and Liu, Deyuan and Fan, Cunhang and Lv, Zhao and Tu, Zhiying and Chu, Dianhui and Sui, Dianbo},
  journal = {arXiv preprint arXiv:2410.07820},
  title   = {Mitigating gender bias in code large language models via model editing},
  year    = {2024},
}

@Article{Liu2024,
  author    = {Liu, Yue and Le-Cong, Thanh and Widyasari, Ratnadira and Tantithamthavorn, Chakkrit and Li, Li and Le, Xuan-Bach D and Lo, David},
  journal   = {ACM Transactions on Software Engineering and Methodology},
  title     = {Refining chatgpt-generated code: Characterizing and mitigating code quality issues},
  year      = {2024},
  number    = {5},
  pages     = {1--26},
  volume    = {33},
  publisher = {ACM New York, NY},
}

@Article{Kojima2022,
  author  = {Kojima, Takeshi and Gu, Shixiang Shane and Reid, Machel and Matsuo, Yutaka and Iwasawa, Yusuke},
  journal = {Advances in neural information processing systems},
  title   = {Large language models are zero-shot reasoners},
  year    = {2022},
  pages   = {22199--22213},
  volume  = {35},
}

@Article{Wang2022,
  author  = {Wang, Xuezhi and Wei, Jason and Schuurmans, Dale and Le, Quoc and Chi, Ed and Narang, Sharan and Chowdhery, Aakanksha and Zhou, Denny},
  journal = {arXiv preprint arXiv:2203.11171},
  title   = {Self-consistency improves chain of thought reasoning in language models},
  year    = {2022},
}

@Article{Yao2023,
  author  = {Yao, Shunyu and Yu, Dian and Zhao, Jeffrey and Shafran, Izhak and Griffiths, Tom and Cao, Yuan and Narasimhan, Karthik},
  journal = {Advances in neural information processing systems},
  title   = {Tree of thoughts: Deliberate problem solving with large language models},
  year    = {2023},
  pages   = {11809--11822},
  volume  = {36},
}

@Article{Abdin2025,
  author  = {Abdin, Marah and Agarwal, Sahaj and Awadallah, Ahmed and Balachandran, Vidhisha and Behl, Harkirat and Chen, Lingjiao and de Rosa, Gustavo and Gunasekar, Suriya and Javaheripi, Mojan and Joshi, Neel and others},
  journal = {arXiv preprint arXiv:2504.21318},
  title   = {Phi-4-reasoning technical report},
  year    = {2025},
}

@Article{Fu2025,
  author    = {Fu, Yujia and Liang, Peng and Tahir, Amjed and Li, Zengyang and Shahin, Mojtaba and Yu, Jiaxin and Chen, Jinfu},
  journal   = {ACM Transactions on Software Engineering and Methodology},
  title     = {Security weaknesses of copilot-generated code in github projects: An empirical study},
  year      = {2025},
  number    = {8},
  pages     = {1--34},
  volume    = {34},
  publisher = {ACM New York, NY},
}

@InProceedings{Majdinasab2024,
  author       = {Majdinasab, Vahid and Bishop, Michael Joshua and Rasheed, Shawn and Moradidakhel, Arghavan and Tahir, Amjed and Khomh, Foutse},
  booktitle    = {2024 IEEE International Conference on Software Analysis, Evolution and Reengineering (SANER)},
  title        = {Assessing the security of github copilot's generated code-a targeted replication study},
  year         = {2024},
  organization = {IEEE},
  pages        = {435--444},
}

@Article{Liu2023a,
  author  = {Liu, Jiawei and Xia, Chunqiu Steven and Wang, Yuyao and Zhang, Lingming},
  journal = {Advances in neural information processing systems},
  title   = {Is your code generated by chatgpt really correct? rigorous evaluation of large language models for code generation},
  year    = {2023},
  pages   = {21558--21572},
  volume  = {36},
}

@Article{Bui2025,
  author    = {Bui, Tuan-Dung and Vu, Thanh Trong and Nguyen, Thu-Trang and Nguyen, Son and Vo, Hieu Dinh},
  journal   = {Journal of Systems and Software},
  title     = {Correctness assessment of code generated by Large Language Models using internal representations},
  year      = {2025},
  pages     = {112570},
  publisher = {Elsevier},
}

@misc{population,
  url = {https://www.bls.gov/cps/},
  year = {2025}
}

@misc{education,
  url = {https://nces.ed.gov/programs/digest/d22/},
  year = {2025}
}

@inproceedings{sun2026multicodeattack,
  title={MultiCodeAttack: Iterative Jailbreak Attacking on LLMs with Multi-Code Prompt Injection},
  author={Sun, Weifeng and Yan, Meng and Yang, Zhou and Chen, Yuchen and Sun, Song and Lo, David},
  booktitle={Findings of the Association for Computational Linguistics: ACL 2026},
  pages={14670--14690},
  year={2026}
}

@article{sun2026cost,
  title={Cost-Effective Adversarial Attacks Against Code LLM with Model Attention},
  author={Sun, Weifeng and Huang, Naiqi and Yan, Meng and Huang, Li and Liu, Zhongxin and Liu, Xiao and Lo, David},
  journal={IEEE Transactions on Software Engineering},
  year={2026},
  publisher={IEEE}
}

\end{document}